\documentstyle[a4,12pt]{article}
\title{\ \\[-7ex] \hrulefill \\ {\bf Renormalisation-theoretic 
analysis of non-equilibrium phase transitions II:  \\ The effect 
of perturbations on rate coefficients in the \BD\ equations} \\[-1ex] }
\author{Jonathan AD Wattis$\dag$ and 
Peter V Coveney$\ddag$\\[-0.5ex] 
{\footnotesize $\dag$Division of Theoretical Mechanics, 
School of Mathematical Sciences,}\\[-0.5ex]
{\footnotesize University of Nottingham,  University Park, Nottingham, 
NG7 2RD, U.K.}\\[-0.5ex]
{\footnotesize $\ddag$Centre for Computational Science, 
Department of Chemistry,}\\[-0.5ex]
{\footnotesize Queen Mary, 
University of London, Mile End Road, London, E1 4NS.}\\[-0.5ex]
{\footnotesize$\dag$ \verb$Jonathan.Wattis@nottingham.ac.uk$ 
\hspace*{8mm}$\ddag$\verb$p.v.coveney@qmul.ac.uk$}\\[-2ex]}
\date{{\footnotesize 3$^{{\rm rd}}$ August, 2001} 
\\[-1ex] \hrulefill}

\newcommand{\beqa}{\begin{eqnarray}}
\newcommand{\eeqa}{\end{eqnarray}}
\newcommand{\beq}{\begin{equation}}
\newcommand{\eeq}{\end{equation}}
\newcommand{\rec}[1]{\mbox{$\frac{1}{#1}$}}
\newcommand{\mfrac}[2]{\mbox{$\frac{#1}{#2}$}}
\newcommand{\pad}[2]{\frac{\partial #1}{\partial #2}}
\newcommand{\padd}[2]{\frac{\partial^2 #1}{\partial #2^2}}
\newcommand{\half}{\mbox{$\frac{1}{2}$}}

\newcommand{\nn}{\nonumber}
\newcommand{\ds}{\displaystyle}
\newcommand{\BD}{Becker-D\"{o}ring}

\let\tilde=\widetilde

\newcommand{\ksum}{\sum_{k=1}^{\infty}}
\newcommand{\ep}{\varepsilon}
\newcommand{\de}{\delta}
\newcommand{\erfc}{{\rm erfc}}

\newcommand{\De}{\Delta}

\newcommand{\Nsum}{\sum_{n=1}^{N-1}}
\newcommand{\Nprod}{\prod_{n=1}^{N-1}}
\newcommand{\prodr}{\prod_{k=1}^{r-1}}
\newcommand{\prodLn}
{\prod_{r=\Lambda_n}^{\!\Lambda_{n\!+\!1}\!-\!1}}
\newcommand{\sumLn}
{\sum_{r=\Lambda_n}^{\!\Lambda_{n\!+\!1}\!-\!1}}

\newcommand{\lbl}[1]{\label{#1}}
\newcommand{\slbl}[1]{\label{#1}}
\begin{document}
\renewcommand{\theequation}
{\arabic{section}.\arabic{equation}}
\newfont{\bbold}{msbm10 scaled\magstep1}
\newcommand{\bbf}[1]{\mbox{{\bbold #1}}}


\maketitle

\vspace*{-15mm}

\begin{abstract}
\noindent We study in detail the application of renormalisation theory to 
models of cluster aggregation and fragmentation of
relevance to nucleation and growth processes.  In particular, 
we investigate the \BD\ equations, originally 
formulated to describe and analyse non-equilibrium phase 
transitions, but more recently generalised to 
describe a wide range of physicochemical problems. 
We consider here rate coefficients 
which depend on the cluster size in a power-law fashion, but now
perturbed by small amplitude random noise. Power-law rate coefficients
arise naturally in the theory of surface-controlled nucleation and
growth processes. The noisy perturbations on these rates reflect 
the effect of microscopic variations in such mean-field coefficients, 
thermal fluctuations and/or experimental uncertainties. 
In the present paper we generalise our earlier work that identified the
nine classes into which all dynamical behaviour must fall (Wattis J A D W
and Coveney P V 2001 {\em J. Phys.A: Math. Gen.}, submitted)
by investigating how random perturbations of the 
rate coefficients influence the steady-state 
and kinetic behaviour of the coarse-grained, renormalised system. We are
hence able to confirm the existence of a set of up to nine 
universality classes for such \BD\ systems. 
\end{abstract} 
\vspace*{-3mm}
\hrulefill

\noindent {\bf PACS}: \\
64.60.-i General studies of phase transitions \\ 
64.60.Ht Dynamic critical phenomena \\ 
82.20.-w Chemical kinetics \\[3ex]
\noindent {\bf Keywords}: \\ 
Nonequilibrium phenomena\\ 
Renormalisation group

\normalsize
\newpage

\section{Introduction} \setcounter{equation}{0}

The purpose of the present paper is to demonstrate the 
robust nature of renormalisation methods in the theoretical
description of 
nucleation and growth processes. We show 
that even in the presence of random perturbations 
the methods presented in our previous papers~\cite{rg,rgpap1} provide correct asymptotic solutions.   
This paper thus extends and completes our earlier published 
work~\cite{rg,rgpap1}, wherein we derived a renormalisation procedure 
for the \BD\ equations and applied it to the case 
where the aggregation and fragmentation rate coefficients 
are of power-law form, given respectively 
by $a_r = a r^p$, $b_{r+1}=b r^p$, $r$ being the 
aggregation number of a cluster (that is, the 
number of monomer particles within it).  
As previously described~\cite{rg,rgpap1}, these coefficients
describe the rate at which a cluster 
of size $r$ adsorbs monomer to grow to size $r+1$, 
and the rate at which a cluster of size $r+1$ sheds 
a monomer; this choice is well known to be appropriate for the description of
surface-limited nucleation and growth processes~\cite{rg,rgpap1}.  
The \BD\ cluster equations are then 
\beq
\dot c_r = J_{r-1} - J_r , \qquad 
J_r = a_r c_r c_1 - b_{r+1} c_{r+1} , 
\eeq
where $c_r(t)$ denotes the concentration of clusters 
of size $r$.  Here we generalise our earlier work to 
consider rates given by the formulae
\beq
a_r = a r^p + \delta_r , 
b_{r+1} = b r^p + \ep_{r+1} . 
\eeq
in which the extra terms represent random perturbations to the 
deterministic rates, and which will be further quantified later on. 
While other authors have considered the effect of noise 
in such systems, the main thrust of such analyses has been 
to elucidate the temporal evolution of perturbations on 
cluster size distribution; see, for example, the work of 
van Dongen \& Ernst~\cite{fluc,fluc2}.

Renormalisation theory has been widely applied 
in the analysis of equilibrium phase transitions in 
statistical physics~\cite{cardy}.   
In statistical mechanics, the basic idea underlying 
renormalisation theory is the transition from a 
microscopic to a macroscopic description of some 
phenomenon by the systematic filtering out of 
unwanted degrees of freedom. 
In equilibrium phase transitions, 
near a critical point the system looks the same on all 
length scales and this physical insight can be 
translated into a set of transformations which
leave the essential physical properties of the system 
unchanged -- a procedure known as renormalisation. 
The widely used term ``The Renormalisation Group'' (RG) is 
technically inappropriate 
since the transformation loses information, and so is 
at most a {\em semi}-group, while 
the procedure comprises many different ideas and
distinct methods, rather than being a formal monolithic 
edifice as the definite article would imply. 
Of more recent interest is the application of 
renormalisation ideas to non-equilibrium phenomena. 
While the physical motivation behind the RG programme of 
coarse-graining microscopic models still seems appropriate 
to obtain the macroscopic properties, the complexity of 
far-from-equilibrium dynamics is such that in practice each specific 
system must be shown to be suitably scale-invariant.

Of central interest in the present paper is the late-time 
asymptotic macroscopic behaviour of complex dynamical 
systems. Bricmont \& Kupiainen have taken ideas from 
renormalisation theory together with asymptotic methods for the 
analysis of diffusive processes including nonlinear parabolic
equations~\cite{bk91,bk92,bk94} while Woodruff has recast multiple 
timescale problems using renormalisation ideas in 
\cite{w93,wm94,w95,w95b}. Woodrfuff's method allows the separation of 
equations for larger-scale phenomena and small-scale 
dynamics from a more general theory.  Velazquez has 
recently used a renormalisation technique in an attempt to 
draw together the theories of Lifshitz-Slyozov coarsening 
and nucleation as modelled by the \BD\ 
equations~ \cite{velaz}.

We apply underlying concepts from renormalisation theory to 
study the \BD\ equations, which were originally formulated to 
study the kinetics of first-order phase transitions. 
They describe the stepwise 
growth and fragmentation of clusters in terms of the rates of 
the individual processes wherein monomer particles join or 
leave each cluster.  The \BD\ equations have recently been 
subjected to more conventional analysis using matched 
asymptotic expansions~\cite{wk98,wk99}. 
Rather than use the coarse-graining approximation, which 
emphasises the discrete nature of the equations, that 
analysis concentrates on the large-time limit where 
continuum approximations become valid~\cite{wk98}.  
We have recently applied generalisations of these equations 
to a wide range of physicochemical processes, ranging from those 
involving surfactant self-assembly~\cite{cw96,cw98} through
RNA-polymer-formation~\cite{ch90,ch95,wc98} to 
cement-setting~\cite{ch95,wc97}. 
In these studies coarse-graining procedures reduce large 
systems of equations down to lower-dimensional--
``mesoscopic''--dynamical systems capable of theoretical 
analysis using standard techniques from the theory of 
differential equations.  The coarse-graining contraction 
procedure summarised below is analogous to other 
renormalization methods used in statistical physics.

There is some similarity between our methods and Woodruff's approach: we
write the microscopic aggregation number $r$ of a 
cluster as $r = (\lambda-1)n+1+k$ where $n$ is of 
mesoscopic size and $k$ is a microscopic 
correction; we then aim to determine the problem on 
the mesoscale in a form  which does not require us to 
simultaneously solve the microscopic problem.  Thus
microscopic detail is filtered out, but we are able to 
construct a simpler model which remains valid on 
larger scales.

Preliminary results of this work were reported in an earlier 
publication~\cite{rg}. In the present paper, the particular and 
physically relevant example of simple 
power-law rate coefficientss is generalised by the addition of 
small random perturbations. These perturbations influence 
the system's steady-states and large-time kinetics and their effects
are studied here in detail. Following a 
brief recapitualtion of the model and 
the coarse-graining scheme underpinning our 
renormalisation procedure in the remainder of the present Section, 
the perturbations are introduced in 
Section~\ref{Gen_Anal}. The central part of  
the present paper is concerned with an analysis of the effects 
which these perturbations have 
on our renormalisation procedure, and how such noise 
influences the contracted description of the model 
(Sections~\ref{nine}). 
The main issue at 
stake is the stability of the models under such minor 
perturbations of rate 
cofficients: our analysis leads directly 
to the identification of a set of nine generic classes of behaviour and 
true {\em universality classes} whose 
asymptotic behaviour is independent of all microscopic 
details (Sections~\ref{nine} \&~\ref{discussion}). We suggest
physicochemical scenarios that may correspond to the generic classes
identified by our renormalsation analysis; future work, especially of
an experimental nature, will be helpful in relating real-world
nucleation and growth processes to these distinct universality classes.

\subsection{The \BD\ cluster equations} 
\lbl{BDs} \setcounter{equation}{0}

In this section we give a basic outline 
of the \BD\ system of equations and their 
properties (consult~\cite{rg,rgpap1} for more details).  
Let us start with a system in which a precursor 
chemical, $P$, spontaneously decays to form the 
monomer $C_1$, at some rate $k_f(p)$ where 
$p=p(t)=[P]$ is the concentration of $P$.  Further, 
we assume that this mechanism is reversible, with 
backard rate $k_b(c_1)$.  The monomer is allowed 
to aggregate, with clusters growing and fragmenting 
according to the usual \BD\ cluster equations.  
Clusters are formed by two processes: either by the 
next smallest cluster size coalescing with a monomer, 
or by the next largest size losing a monomer.  Only 
such monomer--cluster interactions are permitted 
in the \BD\ model of nucleation; cluster--cluster 
interactions are ignored.  The system is thus governed by
\beqa
\dot p & = & k_b(p,c_1) c_1 - k_f(p,c_1) p , \lbl{pdot} \\ 
\dot c_1 & = & k_f(p,c_1) p - k_b(p,c_1) c_1 - J_1 - 
\sum_{r=1}^\infty J_r , \\ 
\dot c_r & = &  J_{r-1} - J_r,\;\;\; (r\geq2), \;\;\;\;\;\;
J_r = a_r c_r c_1 - b_{r+1} c_{r+1}  , \lbl{gdot}
\eeqa
where $c_r(t)$ represents the concentration of 
clusters containing $r$ monomers, and the constants 
$a_r$, $b_r$ are aggregation and fragmentation 
rates respectively.  

In this paper we consider the case for which the 
monomer concentration ($c_1$) is held constant; 
thus the \BD\ equations we are concerned with are
\beq
\dot c_r =  J_{r-1} - J_r,\;\;\; (r\geq2), \;\;\;\;\;\;
J_r = a_r c_r c_1 - b_{r+1} c_{r+1} ,\lbl{fullBD} 
\eeq 
with $c_1$ a given constant; we leave the more general 
formulation (\ref{pdot})-(\ref{gdot}) for a future paper.   
The assumption of a constant monomer concentration 
corresponds to the case where the precursor chemical 
supplies monomer at a rate given by $\dot p = -J_1 - 
\sum_{r=1}^\infty J_r$, so that $\dot c_1=0$.  
This assumption is made in situations where the 
so-called ``pool chemical approximation'' is valid, 
namely where there is a large source of monomer 
species entering into the system at a rate which 
maintains the monomer concentration essentially 
fixed and independent of time. 

Given the rate coefficients $a_r,b_r$ the partition 
function $Q_r$ is defined by $Q_1=1$ and $a_r Q_r 
= b_{r+1} Q_{r+1}$.  This generates the equilibrium 
solution $c_r = Q_r c_1^r$, which is an equilibrium 
solution of the constant monomer model (\ref{fullBD}), 
and also an equilibrium solution of the generalised system 
(\ref{pdot})-(\ref{gdot}) for the particular concentration 
of precursor chemical given by $p = k_b c_1/k_f$. 
In both cases the equilibrium solution corresponds to 
zero flux, that is $J_r=0$ for all $r$, and the flux from 
precursor to monomer is also zero ($k_fp-k_bc_1=0$).  
For certain choices of rate coefficients $a_r,b_r$ and 
certain monomer concentrations $c_1$, the 
equilibrium solution will not decay to zero in the 
limit $r\rightarrow\infty$. In these cases, an alternative 
steady-state solution will be approached in the large-time 
limit. This solution is given by a constant nonzero flux through 
the system, that is $J_r=J$ independent of $r$. This 
condition yields the family of solutions 
\beq
c_r = Q_r c_1^r \left( 1 - J \sum_{k=1}^{r-1} 
\frac{1}{a_k Q_k c_1^{k+1}} \right) , \lbl{gsss}
\eeq
which contains the equilibrium solution as the 
special case $J=0$. 
The steady-state flux $J$ is determined by 
requiring the concentrations $c_r$ to decay to 
zero in the large-$r$ limit, giving 
\beq
J=1\left/\sum_{r=1}^{\infty}\frac{1}{a_rQ_rc_1^{r+1}}\right. .
\lbl{ssflux} \eeq

\subsection{The \BD\ system with power-law coefficients}

We now proceed to consider the renormalisation of \BD\ 
models in which the cluster rate coefficients
are of power-law form, a dependence which is of 
immediate relevance to the description of 
surface-limited aggregation processes. 
We assume the rate coefficients for aggregation and 
fragmentation are respectively 
\beq
a_r=a r^p , \qquad b_{r+1}=b r^p , \lbl{prates}
\eeq 
so that the parameter $\theta=ac_1/b$ is useful for classifying dynamical
behaviour.  The parameter $p$ determines 
the variability of rate with cluster size, with $p>0$ implying 
that large cluster sizes have larger aggregation and 
fragmentation rates, and $p<0$ giving rates which 
decrease with increasing cluster size, the latter case is 
the less physically relevant, but is nevertheless also studied here for the 
sake of completeness.  Typical values for $p$ are
$p=0,\half,\rec{3},\mfrac{2}{3},1$ for the examples of 
linear chain polymerisation, coagulation kinetics in two 
space dimensions, diffusion-limited coagulation in three 
dimensions, surface-limited coagulation in three 
dimensions, and branched chain polymerisation, respectively.  
Since a cluster's volume scales with aggregation number $r$, 
if we assume that clusters are spherical then their surface 
area scales with $r^{2/3}$ and their diameter with $r^{1/3}$, 
accounting for the presence of these exponents. More general 
exponents can be manifest in other situations \cite{costa}.

The partition function $Q_r$ is defined by $Q_r=(a/b)^{r-1}$ 
as in the $p=0$ case considered earlier \cite{wk98}.  
For $\theta\leq1$ the system approaches the equilibrium 
solution given by solving $J_r=0$, that is $c_r=\theta^{r-1}c_1$. 
Note that this solution is independent of $p$, although the way 
in which the equilibrium solution is approached depends on $p$. 
For $\theta>1$ the equilibrium solution diverges at large $r$.
Instead, for $\theta>1$ the system approaches one of a family 
of time-independent solutions in which all fluxes 
are equal; $J_r=J$ for all $r$ implies 
\beq
c_r = \theta^{r-1} c_1 \left( 1 - J \sum_{k=1}^{r-1} 
\frac{1}{b c_1 k^p \theta^k} \right) . 
\eeq
Since, for $\theta>1$ the sum is convergent in the limit 
$r\rightarrow\infty$, the flux which gives 
the least singular behaviour in this limit is 
\beq
J = b c_1 \left/ \ds\ksum k^{-p} \theta^{-k} 
\right. . \lbl{2ssJ}
\eeq

\subsection{Coarse-graining procedure} \lbl{coarse}

Following the general coarse-grained contraction with 
constant mesh size $\lambda$ in aggregation number (so 
that we only retain the aggregation numbers $r=\Lambda_n=
(n-1)\lambda+1$), the kinetic equations reduce to 
\beq
\dot x_n = L_{n-1} - L_n ,\;\;\; (r\geq2), \hspace{9mm}
L_n = \alpha_n x_n x_1^{\lambda} - \beta_{n+1} x_{n+1} 
,\lbl{cgBD} \eeq\beq
\alpha_n=Ta_{\Lambda_n}a_{\Lambda_n+1}\ldots 
a_{\Lambda_{n+1}-1},\hspace{10mm}
\beta_{n+1} =T b_{\Lambda_n+1} b_{\Lambda_n+2} 
\ldots b_{\Lambda_{n+1}} , \lbl{cgrates}
\eeq
where the retained coarse-grained cluster concentrations are relabelled 
as $x_{r}:= c_{r}$ with $x_1 := c_1$ the monomer concentration; $c_1$ 
is not involved in the coarse-graining since it has a special 
role in the \BD\ theory~\cite{cw96,rg}. The parameters 
$\alpha_n,\beta_n$ are the coarse-grained 
aggregation and fragmentation rates, now representing 
the addition or removal of $\lambda$ monomers to or from 
a cluster (rather than just a single monomer which 
occurs in the full \BD\ system).  This flux of matter is 
denoted by $L_n$.   The concentration $x_n(t)$ is 
representative of the concentrations $c_r$ for 
cluster sizes $(\Lambda_{n-1}+1) \leq r \leq \Lambda_n$. 
The factor $T$ represents a change of timescale which 
ensures that the large time asymptotic behaviour of 
the reduced system coincides exactly with the original 
fine-grained system in the case of size-independent 
aggregation and fragmentation rates ($a_r=a$, $b_r=b$). Technically 
speaking, the need to redefine the timescale makes this 
procedure a {\em dynamical} renormalisation.

\subsection{Coarse-graining of power law coefficients}
\slbl{sec1p4}

If the rate coefficients in the original formulation in 
eqns~(\ref{fullBD}) 
are determined by simple power laws, namely $a_r = a r^p$ 
and $b_{r+1} = b r^p$, then the coefficients in the reduced 
model are 
\beq
\alpha_n = a^\lambda \left\{ [(r-1)\lambda+1] 
[(n-1)\lambda+2] \ldots [n\lambda]  \right\}^p . 
\lbl{1p9} \eeq
Thus 
\beqa
\log \alpha_n & = & \nn
\lambda \log a + p \sum_{j=1}^\lambda 
\log(n\lambda-\lambda+j) \;\; \approx \;\; 
\lambda\log a + p \int_0^\lambda 
\log(n\lambda-\lambda+x)dx \\
& = & \lambda\log a + p \lambda \,\left[\, 
\log(n\lambda) - 1 + (1\!-\!n) \log \left( 1 - 
\rec{n} \right) \right] . \lbl{rgg1app}
\eeqa
For large $n$ this asymptotes to $\log\alpha_n\sim 
\lambda \log a + p\lambda \log(\lambda n)$, 
so for simplicity we shall take $\alpha_n = 
(a \lambda^p n^p)^\lambda$, which differs slightly at small 
values of $n$.  The backward rate coefficient is then 
$\beta_{n+1} = (b\lambda^p n^p)^\lambda$.

The new system has its own $\theta$ parameter determining 
the balance between aggregation and fragmentation 
rates in the system which, for the moment, we shall call 
$\tilde{\theta} = \alpha_r x_1^\lambda 
/ \beta_{r+1} = \theta^\lambda$, so contraction of the 
system maps $\theta$ to $\theta^\lambda$.   The parameter 
$\theta$ thus plays an important role in the renormalisation 
procedure, the fixed points of this mapping corresponding to 
$\theta=0,1,\infty$; hence such systems are of 
special interest to us.  Also the contraction maps coefficients 
with exponent $p$ to those with exponent $p\lambda$. Thus 
following a contraction, there are only three limits to consider: 
small $p$ (namely $p=0$) and large $p$ (positive and negative).

The effect of coarse-graining a \BD\ system is to modify the 
rate coefficients, by the map $(\theta,p)\mapsto 
(\theta^\lambda,\lambda p)$.  If $\lambda$ 
is allowed to take on large values, there are only nine 
combinations of $(\theta,p)$ which merit attention, namely 
all possible combinations of $\theta=\{0,1,\infty\}$ and 
$p=0,p>0,p<0$.  These nine cases and their associated fixed 
points will be the basis of the analysis presented 
in Section 3 of this paper, which follows on from  
Section~\ref{Gen_Anal} where we develop the general theory for perturbed rates 
for the coarse-grained \BD\ system with arbitrary $\lambda$.

\section{General analysis of the role of noise in coarse-grained \BD\ systems}
\lbl{Gen_Anal}
\setcounter{equation}{0}

\subsection{Form of rate perturbations}
\lbl{noise-def-sec}

Whilst a simple power-law description of rate coefficients 
might give the correct 
general behaviour over a large range of 
aggregation numbers, in any given physical system 
there will be some minor deviations from such a regular descriptions 
which may be due to a number of separate effects. For instance, the 
presence of microscopic fluctuations and/or experimental uncertainties suggest that one should be concerned about the stability of such theoretical models to minor modifications in the assumed rate coefficients. Indeed, Figure 2.3 
of Lewis's article in the 
monograph~\cite{lewis} shows that the dependence 
of free energy on cluster size can in reality be much more 
complicated than the simple smooth curves obtained 
from simple derivations (for example by considering 
surface and bulk energies and assuming all clusters are 
spheres, discs or needles; or by working back from 
simple laws for aggregation and fragmentation rates). In mathematical 
terms, therefore,
we wish to know whether small random perturbations to the rates alter the 
results we have derived previously using renormalisation methods. The 
hope is that they will not have a major effect, so that the renormalisation 
scheme is stable with respect to such perturbations and points to the existence of true universality classes.  To analyse the effect of 
such perturbations, we perturb the rates from a simple power 
law form by small random amounts, in such a way that all 
rates remain positive.  Thus we assume that the  
rate coefficients have the form 
\beq
a_r = a r^p + \de_r , \hspace*{9mm}
b_{r+1} = b r^p + \ep_{r+1} , \lbl{noise}
\eeq
where $\de_r,\ep_r$ are independent random variables, with 
small mean and small variance, so we can assume 
that $\de_r\sim\ep_r\sim\nu\ll1$ for all $r$. 
Many of the approximations used in the ensuing 
analysis rely on the noise being small ($\nu\ll1$). 
Even with such a restrictive assumption, by 
calculating the first higher order term in the 
asymptotic expansion, many important and 
interesting results are obtained.  Later on, we shall discuss 
cases in which $\nu\gg1$ and $\nu={\cal O}(1)$.  In the nine 
special cases analysed below, the constants $a,b$ in eqns~(\ref{noise}) 
will be taken to be either zero or unity. When 
$\lim_{r\rightarrow\infty}a_r$ or $\lim_{r\rightarrow\infty}b_r$ 
is zero, the corresponding perturbation ($\delta_r,\ep_r$ 
respectively) will be assumed to be strictly positive and have 
mean $\nu$.  Specifically, we shall assume that they 
are distributed according to some continuous probability 
distribution functions $f_\de(x)$, $f_\ep(x)$ respectively,  
with $f(\cdot)=0$ whenever $\de,\ep$ lies more than 
${\cal O}(\nu)$ away from zero.  The expectation operator 
${\bbf E}[\cdot]$ is defined by
\beq
{\bbf E}[g(\de)] = \int g(x) f_\de(x) dx , \qquad {\rm and} 
\qquad {\bbf E}[g(\ep)] = \int g(x) f_\ep(x) dx , 
\eeq 
the integrals being over all real $x$.  The variance 
operator ${\bbf V}[\cdot]$ is defined by 
\beq 
{\bbf V}[g] = {\bbf E}[g^2] - {\bbf E}[g]^2 . 
\eeq 
The conditions 
on the mean reduces to ${\bbf E}[\de_r]=\nu$ 
and/or ${\bbf E}[\ep_r]=\nu$ for all $r$; in cases 
where $\lim_{r\rightarrow\infty}a_r$ and/or 
$\lim_{r\rightarrow\infty}b_r$ are unity or larger $\de_r,\ep_r$ 
will be allowed to take negative values as well as positive 
and we assume that they have zero mean and their 
variance is ${\cal O}(\nu^2)$, (e.g. 
${\bbf E}[\de_r]=0$ with ${\bbf E}[\de_r^2]={\cal O}(\nu^2)$). 
In all cases the perturbations are assumed to be small enough 
that the total rate constants in equation (\ref{noise}) are all positive. 
 
\subsection{Effect of noise on cluster partition functions}
\slbl{effNQ-sec}

We first derive and analyse the modified equilibrium 
and steady states which are approached in the large-time 
limit. The kinetics by which these states are achieved is 
described in detail later (Section \ref{effNkin-sec}).  
First we aim to find the effect which the perturbations to the 
rate coefficients ($\de_r,\ep_r$) have on the cluster partition 
function $Q_r$ in each of the nine special cases described 
above. Since the partition function is defined in terms of the rate 
constants, the presence of non-zero perturbations 
$\de_r,\ep_r$ will affect $Q_r$. Due to the definition 
$c_1=1$, the partition function is identical to the equilibrium 
configuration. However, in Cases III, VI, IX and V (if $p>1$) it is 
not the equilibrium solution which is approached in the 
large-time limit; instead the system evolves to a steady-state 
solution.   Perturbations to the rate coefficients ($\de_r,\ep_r$) 
thus modify the steady-state and equilibrium solutions through 
the partition function, and such modifications will be found here.  
In general the partition function is given by 
\beq
Q_r = \prod_{k=1}^{r-1} \frac{ a_k }{ b_{k+1} } , 
\eeq
whilst steady-state solutions are determined by the constant 
flux condition $J=J_r=a_r c_r - b_{r+1} c_{r+1}$, for 
all values of $r$. We shall pay particular attention to the large aggregation number ($r$) 
behaviour, since this determines the observed behaviour 
of the system on the mesoscopic scale, and is the region we 
shall probe later with the coarse-grained contraction of this 
perturbed \BD\ system.

\subsection{Effect of noise on cluster growth kinetics}
\slbl{effNkin-sec}

The equations we study in this section are the kinetic 
equations with noisy coefficients, namely 
\beq
\dot c_r = (a(r-1)^p+\de_{r-1})c_{r-1} - (ar^p+\de_r)c_r - 
(b(r-1)^p+\ep_r)c_r + (br^p+\ep_{r+1})c_{r+1},\qquad(r\geq2). 
\lbl{nkineq} \eeq
Having found the states which are approached in the 
large-time limit, we now find time-dependent solutions by 
using large-time asymptotics.  We write the solution as 
$c_r(t)=Q_r\psi(r,t)$ or $c_r(t)=c_r^{{\rm sss}} \psi(r,t)$ in 
order to investigate the manner in which 
$\psi(r,t)\rightarrow1$ as $t\rightarrow+\infty$. In general, 
this leads to 
\beq
\dot \psi_r = a_r (\psi_{r+1} - \psi_r ) - b_r (\psi_r - \psi_{r-1} ) - 
\frac{J}{c_r^{{\rm sss}}} (\psi_{r+1}-\psi_{r-1}) , 
\eeq
where $J$ represents the steady-state flux into the 
system; convergence to equilibrium corresponds to $J=0$.

\subsection{Effect of perturbations on rate coefficients in the 
contracted system}
\slbl{effNrates-sec}

The rate coefficients following contraction are given by 
\beq
\alpha_n=T\prod_{r=\Lambda_n}^{\Lambda_{n+1}-1} a_r ,  
\hspace*{9mm} 
\beta_{n+1}=T\prod_{r=\Lambda_n}^{\Lambda_{n+1}-1}b_{r+1}.
\eeq
where we shall use $T = \lambda^{-p\lambda}$ so that 
the `clean' coefficients $a_r = a r^p$, $b_{r+1} = b r^p$ are 
mapped to $\alpha_n = \alpha n^{p\lambda}$, $\beta_{n+1} = 
\beta n^{p\lambda}$ with $\alpha = a^\lambda$ and $\beta 
= b^\lambda$ (see Section \ref{sec1p4} and \cite{rgpap1}). 
Since the nine special cases which will be studied in detail later 
correspond to $a,b$ equal to zero or unity, following the 
coarse-graining contraction we are concerned with establishing 
like-for-like correspondence in which $\alpha,\beta$ 
equal zero or unity.   
Here $T$ is chosen to simplify the algebra; an alternative 
expression could be used, with a consequent increase in the 
complexity of the ensuing equations.  Alternative choices for $T$ 
merely influence the units of time, and have no effect on the 
equilibrium or steady-state solutions or the effect of the rate 
perturbations; the large-time asymptotics will only be affected 
by a linear transformation to the time variable, $t$. 
The problems we are concerned with here involve 
properties of the noise following contraction, for example  
whether its amplitude is dependent on aggregation number, 
and its order of magnitude as a function of $\nu$.  
We now reduce the randomly perturbed coefficients from 
(\ref{noise}) in the microscopic description of 
cluster-formation to the mesoscopic description in which 
$\alpha_n = \alpha n^{p\lambda}+\De_n$, 
$\beta_{n+1} = \beta n^{p\lambda} + E_{n+1}$ with 
the aim of finding how the perturbations $\De_n,E_n$ in the 
contracted model depend on the perturbations $\de_r,\ep_r$ 
in the full description of the model.  In some cases this is trivial, 
since $a=0$ or $b=0$, and in these cases $\De_n,E_{n+1}$ is 
the product of a set of $\de_r,\ep_r$, but in other cases the 
relationship is more complex, and in such cases we give the 
leading-order expression for $\De_n,E_{n+1}$ in the 
asymptotic limit $\nu\rightarrow0$.    We have
\beq
\alpha_n = 
T \left( \prod_{r=\Lambda_n}^{\Lambda_{n+1}-1} a r^p \right) 
\left( \prod_{r=\Lambda_n}^{\Lambda_{n+1}-1} 
\left[1 + \frac{\delta_r}{a r^p} \right] \right) \sim 
T \left( \prod_{r=\Lambda_n}^{\Lambda_{n+1}-1} a r^p \right) 
\left( 1 + \sum_{r=\Lambda_n}^{\Lambda_{n+1}-1} 
\frac{\delta_r}{a r^p} \right) , 
\lbl{2p16a} \eeq
since $\delta_r = {\cal O}(\nu) \ll a r^p = {\cal O}(1)$; 
correction terms to the above approximation are thus 
${\cal O}(\nu^2)$.  
Using (\ref{1p9})--(\ref{rgg1app}), the leading-order term 
is approximated by $\alpha n^{p\lambda}$ where 
$\alpha = T a^\lambda \lambda^{p\lambda}$ and the 
first correction term due to the rate perturbations is defined by 
\beq
\De_n = \alpha n^{p\lambda} 
\sum_{r=\Lambda_n}^{\Lambda_{n+1}-1} 
\frac{\delta_r}{a r^p}  .
\eeq
Similarly for the fragmentation rates we have 
$\beta_{n+1}=\beta n^{p\lambda} + E_{n+1}$ with 
$\beta=T b^\lambda \lambda^{p\lambda} $ and 
\beq
E_{n+1} = \beta n^{p\lambda} 
\sum_{r=\Lambda_n}^{\Lambda_{n+1}-1} 
\frac{\ep_{r+1}}{b r^p} .
\lbl{2p16c} \eeq

\subsection{Consistency and accuracy analysis}
\slbl{effNcQ-sec}
 
Having derived coarse-grained rate coefficients 
which include the leading order corrections due to 
random perturbations from a power law, we now 
use these rate coefficients to construct a partition 
function $\Upsilon_n$ for the coarse-grained model 
from the macroscopic rates $\alpha_n,\beta_n$.  
This is determined by 
\beq
\Upsilon_1=1 , \hspace*{9mm} 
\alpha_n \Upsilon_n = \beta_{n+1} \Upsilon_{n+1} , 
\qquad (n\geq1) , 
\eeq
which implies 
\beq
\Upsilon_N = \Nprod \frac{\alpha_n}{\beta_{n+1}} 
\qquad {\rm and} \quad  \log \Upsilon_N = 
\sum_{n=1}^{N-1} ( \log \alpha_n - \log \beta_{n+1}) .
\lbl{URdef} \eeq
We can then compare this result with the full partition 
function as a check on the coarse-graining procedures, 
with the hope of finding $\Upsilon_N=Q_R$ when $R = 
\Lambda_N$ to leading order in $\nu$. For cases in which 
a steady-state solution is approached as $t\rightarrow\infty$, 
we compare the steady-state solution $x_n^{{\rm sss}}$ 
calculated from the noisy contracted rates, 
$\alpha_n,\beta_{n+1}$, with the steady-state solution 
of the full model, $c_r^{{\rm sss}}$.  If moreover $c_r=x_n$ when 
$r=\Lambda_n$ then we will have demonstrated that no 
crucial information has been lost in the coarse-graining 
reduction.  This test thus amounts to a consistency check 
which we shall carry out not just at leading order, but also 
to first order in $\nu$ (that is in $\De_r,E_{r+1}$) for the purposes of 
accuracy assessment. These checks will enable us to 
demonstrate if and when we have successfully bridged the 
scales from microscopic to mesoscopic for certain 
choices of the coarse-graining parameter $\lambda$.

\subsection{Effect of noise on the late-time kinetics of the contracted 
system} \slbl{effNckin-sec}

Having checked that the equilibrium and steady-state 
solutions of the microscopic description of the system 
have been faithfully reproduced in the coarse-grained 
system, we perform one final calculation to check that 
the late-time asymptotic form of the kinetics is also correctly 
replicated.  In general the contracted system of equations 
can be written using (\ref{cgBD}) and (\ref{2p16a})--(\ref{2p16c}) as 
\beq   \dot x_n = 
\,\left[\, \alpha (n\!-\!1)^{p\lambda}+\De_{n-1}\,\right]\,x_{n-1}- 
\,\left[\, \beta (n\!-\!1)^{p\lambda} + E_n \,\right]\, x_n - 
\,\left[\, \alpha n^{p\lambda} + \De_n \,\right]\, x_n + 
\,\left[\, \beta n^{p\lambda} + E_{n+1} \,\right]\, x_{n+1} . 
\eeq
In the following section we examine the equilibrium and 
steady-state solutions of this model, as well as the manner 
in which the time-dependent solutions convergence to these 
solutions.  We then compare this behaviour with the properties 
of the microscopic system prior to coarse-graining.   
We now analyse in greater detail the nine cases summarised 
at the end of section \ref{sec1p4}; that is those corresponding 
to $p=0$, $p>0$ and $p<0$ in the three cases $\theta=0,1,\infty$. 
The conditions $\theta=0,1,\infty$ correspond to $(\alpha,\beta)$
equalling $(0,1)$, $(1,1)$, $(1,0)$ respectively.  
The fourth case $(\alpha,\beta)=(0,0)$ yields a special limitting 
system which will be described in Section \ref{CaseX}.

\section{The nine generic classes of asymptotic behaviour}
\lbl{nine}
\setcounter{equation}{0}

Nine distinct generic classes of asymptotic behaviour were picked out 
by the RG analysis in Section 1.4 (see also~\cite{rg,rgpap1}).For each 
class or case we shall perform the five calculations 
described above. We start with the effect of the perturbations 
on the partition function, and hence the equilibrium solution. 
In cases where a steady-state is approached rather than the 
equilibrium we also calculate the modified form of the steady-state
solution.  We then calculate the effect on the kinetics of approach 
to steady-state or equilibrium.  Following these two calculations 
we turn to the contracted description of the system.  
The latter three calculations combine the problems 
of dealing with the noise and the coarse-grained system. 
We analyse the contracted \BD\ system derived from the 
full microscopic model with rate coefficients given by a 
combination of a power-law with random perturbations,  and 
investigate whether coarse-graining averages noise 
out of the system or exaggerates its effect.  In so doing, we discover how 
perturbations to the microscopic rates manifest themselves in the rates of 
the contracted system.  We then establish consistency by showing 
that the partition function calculated from the coarse-grained 
rates is the same, to leading order, as the partition function from 
the microscopic system sampled over the coarse mesh.  Finally, 
we analyse the effect of the perturbations in the mesoscopic 
model on the large-time kinetics of the coarse-grained system.

\subsection{Case I}

This case corresponds to a fragmentation-dominated 
system where the fragmentation rate is independent of 
cluster size, that is the rate at which large 
clusters shed monomers is the same as that for small clusters. 
Such situations may arise in certain types of polymer degradation, 
if the breakdown of chains takes place through end-monomer fission processes.


In the non-perturbed version of this case, $a_r=0$ for all $r$, 
so the partition function is identically zero.  Introducing 
noise to the aggregation coefficient makes the partition 
function non-zero
\beq
Q_r = \prodr \frac{\de_k}{1+\ep_{k+1}} \sim 
\left( \prodr \de_k \right) \left( 1 - 
\sum_{k=1}^{r-1} \ep_{k+1} + {\cal O}(\nu^2) \right) . 
\lbl{eNQI} \eeq
Thus the partition function $Q_r$ depends primarily on the 
perturbations $\delta_r$; the perturbations $\ep_r$ only 
enter at higher order.  Since ${\bbf E}[\de_r]=\nu\ll1$, 
$Q_r$ decays rapidly to zero with increasing $r$, according 
to $Q_r={\cal O}(\nu^{r-1})$, thus $Q_{r+1}\ll Q_r$ for all $r$. 
Case I may hence correspond to a situation in which 
finite-length polymers or oligomers break down via stepwise 
fission processes.


In the noisy case, there is a non-zero partition function 
$Q_r$ to which the system converges. To determine the 
manner of this convergence we transform from $c_r(t)$ 
to new coordinates, $c_r=Q_r\psi_r(t)$ so that at equilibrium 
$\psi_r\equiv1$.  Then $\psi_r(t)$ satisfies 
\beq
\dot \psi_r = \psi_{r-1} - \psi_r + \de_r(\psi_{r+1}-\psi_r) - 
\ep_r(\psi_r-\psi_{r-1}) ; 
\eeq
for large aggregation numbers ($r$) and at large 
times ($t$), the continuum limit is valid and 
this equation goes over into the following partial differential equation: 
\beq
\pad{\psi}{t} = \frac{1}{2} \padd{\psi}{r} \left(1+\de_r+\ep_r 
\right) - \pad{\psi}{r} \left( 1 +\ep_r-\de_r \right) . 
\lbl{421pde} \eeq
The solutions of this type of equation typically approach equilibrium via 
a diffusive wave, whose position we shall denote by $r=s(t)$.  
We transform from $r$ to the new position variable $z=r-s(t)$ 
relative to the wavefront.  The leading order terms are then 
those involving $\pad{\psi}{z}$, formally yielding the equation 
$\dot s = 1+\ep_s-\de_s$ for the position of the wave.  
The expected value of the first perturbation (due to $\de_r$) is zero, since  
${\bbf E}[\ep_r]=0$ for all $r$, but that of the second (due to $\ep_r$) is 
positive, that is ${\bbf E}[\de_r]=\nu$ for all $r$. Thus we have 
${\bbf E}[\dot s]=1-\nu$, and the presence of perturbations 
slows the wave. 
This is not, however, a leading order effect since small noise 
makes only a small difference to the system's approach to 
equilibrium. Higher order terms from equation (\ref{421pde}) lead 
to a description of the shape of the wavefront.   If we start 
the system from a state of compact support 
(that is $c_r(0)=0$ for $r\geq R$ for some $R<\infty$),  
then the large-time and large aggregation number 
asymptotic solution is given by 
\beq
c_r(t) = \half \, Q_r \, \erfc\left( 
\frac{r-(1\!-\!\nu)t}{\sqrt{2\,(1\!+\!\nu)\,t}} \right) , 
\lbl{eNkin1sol} \eeq
where $Q_r$ is the partition function defined by 
equation (\ref{eNQI}).  
Thus as well as travelling slightly more slowly than in the 
noiseless case, the wave is widened slightly, being spread 
out over a larger number of cluster sizes.  These 
effects would be hard to measure experimentally 
since the typical concentration of large cluster sizes 
would be extremely small.  Both these effects are due to the 
differing expected values of the perturbations to the 
forward and backward rate constants (namely that 
${\bbf E}[\ep_r]=0$ whilst ${\bbf E}[\de_r]=\nu$).


Moving now to the coarse-grained, mesoscopic description 
of the problem, we find $a_r=0$, $b_r=1$ implies 
$\alpha_n = \De_n$ and  $\beta_{n+1} = 1+E_{n+1}$ where
\beq
\De_n = \prodLn \de_r , \hspace*{9mm} 
E_{n+1} = \sumLn \ep_{r+1} + {\cal O}(\nu^2) . 
\lbl{eNabI} \eeq
Thus following an application of the coarse-graining contraction 
procedure, to ${\cal O}(1)$ the rates are $\alpha_n=0$ 
and $\beta_{n+1}=1$ with small perturbations to the rates, 
$\Delta_r = {\cal O}(\nu^\lambda)$ and $E_{r+1} = {\cal O}
(\nu)$. Thus the perturbations to the aggregation 
rates are much smaller than those to the fragmentation rates, 
making the system {\em appear} more fragmentation-dominated 
than the uncontracted version. However, this effect is due to 
the contracted model only representing some cluster sizes: 
growth from $x_r$ to $x_{r+1}$ requires the addition of 
$\lambda$ monomers and hence happens at a rate of ${\cal O} 
(\nu^\lambda)$ rather than the ${\cal O}(\nu)$ rate for 
growth from $c_r$ to $c_{r+1}$ in the full model.  Thus the 
contracted description is dominated by fragmentation to 
the correct extent. 
Since the perturbations in the full model are independent, 
and distributed with ${\bbf E}[\de_r]=\nu$, ${\bbf E}[\ep_r]=0$, 
${\bbf V}[\de_r]={\cal O}(\nu^2)$, 
${\bbf V}[\ep_r]={\cal O}(\nu^2)$, in the contracted 
description we have ${\bbf E}[\De_n]=\nu^\lambda$, 
${\bbf E}[E_n]=0$, ${\bbf V}[\De_n]={\cal O}(\nu^{2\lambda})$, 
and ${\bbf V}[E_n]={\cal O}(\nu^2)$. 


We now use the contracted rates given by $\alpha_n=
\Delta_n$ and $\beta_{n+1} = 1 + E_{n+1}$ to calculate 
a partition function for the coarse-grained system. Using 
(\ref{URdef}) and (\ref{eNabI}), we find
\beq
\Upsilon_N = \left( \Nprod \De_n \right) 
\left( 1 - \Nsum E_{n+1} + {\cal O}(\nu^2) \right) 
= \left( \prod_{r=1}^{\Lambda_N-1} \de_r \right) 
\left( 1- \sum_{r=1}^{\Lambda_N-1} \ep_{r+1} + 
{\cal O}(\nu^2) \right) . \lbl{eNcQI} 
\eeq
Thus the first two terms of the equilibrium solution of the 
reduced model agree exactly with that of $Q_R$ with 
$R=\Lambda_N$ in the full model (\ref{eNQI}).


In this case we know that the system will tend to the 
equilibrium solution given by $x_n=\Upsilon_n$ as in 
(\ref{eNcQI}), which closely approximates $Q_r$ 
(\ref{eNcQI}) with $r = \Lambda_n$.  To find the way in 
which the solution is approached, we use a similar 
method to that successfully applied to the miroscopic 
description above, namely that of transforming to new 
dependent variables $\psi_n(t)$ given by $x_n(t)=
\Upsilon_n\psi_n(t)$. Then to first order in $\nu$ we have 
\beq
\dot \psi_n = \psi_{n-1} - \psi_n - E_n(\psi_n-\psi_{n-1}) , 
\eeq 
since in this case $\De_n = {\cal O}(\nu^\lambda)$ and 
$E_n = {\cal O}(\nu)$.   For large times and large $n$-values 
the continuum approximation 
\beq
\pad{\psi}{t} = \half(1+E_n)\padd{\psi}{n} - 
(1+E_n)\pad{\psi}{n} , 
\eeq
is formally valid.  Since ${\bbf E}[E_n] = 0$, the diffusive wave 
travels at unit expected speed and suffers no ${\cal O}(\nu)$ 
correction term.  Thus if the system is initiated from $x_n(0)=0$ 
for $n\geq2$, the equilibrium is reached is via a diffusive 
wave moving from $n=1$ to large $n$ leaving the new 
equilibrium solution behind it as described by $x_n(t) \sim 
\half\Upsilon_n\erfc((n\!-\!t)/\sqrt{2t})$.  The noise thus has 
no effect at leading order, or at ${\cal O}(\nu)$, although 
higher order terms will influence the evolution of the system. 
Thus, in this case, coarse-graining has {\em reduced} the effect of 
the noise, since in the non-coarse-grained case, the kinetics 
{\em are} affected by ${\cal O}(\nu)$ terms, decelerating 
and widening the wavefront as described by 
equation~(\ref{eNkin1sol}).
 
\subsection{Case II}

In this case the aggregation and fragmentation rates 
are both relevant and both size independent. This situation 
can be expected to arise in a wide variety of nucleation and growth 
problems, for example it is likely to pertain for cluster 
formation at saturation or low supersaturation levels during 
crystal growth.


In this case introducing perturbations $\delta_r$, $\ep_r$ 
to the rates $a_r=ar^p$, $b_{r+1}=b r^p$  as in (\ref{noise}) 
modifies the partition function from $Q_r=1$ for all $r$ to 
\beq
Q_r \sim 1 + \sum_{k=1}^{r-1} (\de_k-\ep_{k+1}) + 
{\cal O}(\nu^2) . \lbl{eNQII} 
\eeq
Small amplitude noise in the coefficients thus does not 
affect the leading-order behaviour of the system, at small 
values of $r$.  To examine the large-$r$ behaviour, we use 
the central limit theorem.  For simplicity we 
assume that each of the random variables $\de_k,\ep_k$ 
is distributed uniformly on the interval $[-\nu,\nu]$, thus 
each has an expected value of zero and variance of 
$\sigma^2=\rec{3}\nu^2$. The difference $\de_k-\ep_{k+1}$ 
thus has mean of zero and variance of $\mfrac{2}{3}\nu^2$; and 
the above sum (equivalent to $Q_r-1$) has mean zero and 
variance $\mfrac{2}{3}(r-1)\nu^2$.  At large values of $r$, 
the central limit theorem implies the sum can be approximated 
by a normally distributed random variable with zero mean and 
variance $\sigma^2=\mfrac{2}{3}r\nu^2$. Thus perturbations 
have a cumulative effect and may become significant when 
$r={\cal O}(\nu^{-2})$; at this order of magnitude, 
the approximation in equation (\ref{eNQII}) ceases to be valid, since 
neglected higher order terms then become significant.


For Case II the kinetic equations are
\beq
\dot c_r = (1+\de_{r-1}) c_{r-1} - (1+\ep_r) c_r - 
(1+\de_r) c_r + (1+\ep_{r+1}) c_{r+1} , 
\eeq
where $\de_r,\ep_r$ can be positive of negative, and are 
${\cal O}(\nu)$ with ${\bbf E}[\de_r]={\bbf E}[\ep_r]=0$. 
Formally, for large-times and large aggregation numbers, 
this leads to the continuum equation 
\beq
\pad{\psi}{t}=\left(1+\half(\de_r+\ep_r)\right)\padd{\psi}{r}
+(\de_r-\ep_r)\pad{\psi}{r},
\eeq
for $\psi(r,t)=c_r(t)/Q_r$. Since 
${\bbf E}[\de_r-\ep_r]=0$, and ${\bbf E}[(\de_r-\ep_r)^2] 
\sim\nu^2 \ll 1$, diffusion dominates the advection terms, 
and there is no overall driving force on the diffusive 
wave. Effectively, it is pinned at $r=1$ and equilibrium 
is reached by purely diffusive mechanisms. If we use 
compact initial conditions (that is $c_r(0)=0$ for all $r\geq R$ 
for some $R<\infty$) then the large-time asymptotic 
solution is 
\beq 
c_r(t) = Q_r \erfc \left( \frac{r}{2\sqrt{t}}\right) . 
\lbl{IImicroasysol} \eeq 
Note that 
${\bbf E}[\de_r]=0={\bbf E}[\ep_r]$ implies that the 
width scale is unchanged by the perturbations, 
in contrast with Case I, where the wave was widened 
due to ${\bbf E}[\delta_r]>0$.


Following the coarse-graining contraction, the reaction 
rates are given by 
$\alpha_n=1+\De_n$ and $\beta_{n+1}=1+E_{n+1}$, where 
\beq
\De_n = \sumLn \de_r + {\cal O}(\nu^2), \hspace*{9mm} 
E_{n+1} = \sumLn \ep_{r+1} + {\cal O}(\nu^2) . 
\lbl{eNabII} \eeq
Thus a system in which aggregation and fragmentation 
are balanced is mapped to a similar system following a 
coarse-grained reduction. Since ${\bbf E}[\de_r] = 0 = 
{\bbf E}[\ep_r]$ and ${\bbf V}[\de_r],{\bbf V}[\ep_r] = 
{\cal O}(\nu^2)$ we have ${\bbf E}[\De_n] = 0 = {\bbf E}[E_n]$ 
and ${\bbf V}[\De_n],{\bbf V}[E_n] = {\cal O}(\nu)$.  
The central limit theorem implies that the variances of the 
noise will increase linearly with $\lambda$.  Since both 
$\delta_k,\ep_k$ have variance proportional to $\nu^2$,  
$\De_r,E_r$ have  variance proportional to $\lambda\nu^2$.   
So in order for the noise to remain a small correction term, 
one has to ensure that $\lambda\ll\nu^{-2}$; for small 
perturbations ($\nu\ll1$) this does not constitute a 
significant restriction.


Using (\ref{URdef}) and (\ref{eNabII}), we find the 
partition function $\Upsilon_n$ for the coarse-grained 
system is given by 
\beq
\Upsilon_N=1+\Nsum(\De_n-E_{n+1})+{\cal O}(\nu^2)=1+
\sum_{r=1}^{\Lambda_N-1}(\de_r-\ep_{r+1})+{\cal O}(\nu^2). 
\eeq
Thus the first two terms of the equilibrium solution of the 
reduced model agree exactly with that of $Q_R$ with 
$R=\Lambda_N$ in the full model (\ref{eNQII}). Thus in the 
presence of noise, the contracted system tends to an 
equilibrium solution of the same form, to first order, as the 
full system.


The kinetic equations determining the approach to 
equilibrium are
\beq
\dot x_n = (1+\Delta_{n-1}) x_{n-1} - (1+E_n) x_n - 
(1+\Delta_n) x_n + (1+E_{n+1}) x_{n+1} , 
\eeq
where $\De_n,E_n$ are ${\cal O}(\nu)$ with 
${\bbf E}[\De_n]={\bbf E}[E_n]=0$.  The system tends to 
the equilibrium solution $x_n=\Upsilon_n$ by a purely 
diffusive mechanism as described formally by the 
continuum limit equation 
\beq
\pad{\psi}{t}=\left(1+\frac{\De_n+E_n}{2}\right) \padd{\psi}{n}
+ \left( \De_n - E_n \right) \pad{\psi}{n} 
\eeq 
for $\psi(n,t)=x_n(t)/\Upsilon_n$, so that $\psi\rightarrow1$ 
as $t\rightarrow\infty$.  Since the expected values of 
perturbations $\De_n$ and $E_n$ are both zero, there are no 
${\cal O}(\nu)$ correction terms to $\psi(n,t)$ in the large-time 
limit, and we have the solution $x_n(t)=\Upsilon_n \erfc( n / 2 
\sqrt{t})$. Thus to ${\cal O}(\nu)$ the large-time kinetics of 
the system are almost identical to the uncontracted model. 
Using $r=\Lambda_n$, $c_r=x_n$ and $Q_r=\Upsilon_n$, 
the above formula for $x_n(t)$ implies $c_r = Q_r \erfc ( 
r/2\lambda\sqrt{t})$ whereas analysis of the full model gave 
(\ref{IImicroasysol}); thus the only difference between full and 
contracted models is in the timescale.

\subsection{Case III}

This case models situations which favour the formation 
of clusters since the system is dominated by aggregation; however 
the rate of aggregation is size-independent so that 
nucleation and cluster growth processes are balanced. One 
physicochemical scenario that this Case may describe arise in certain surfactant self-assembly processes, in which amphiphile monomers attach to a growing assembly (for example, wormlike micelles, vesicles, and so on).


In the pure aggregation case with no noise the partition 
function is not defined since all the fragmentation 
coefficients $b_r$ vanish.  The presence of noise makes it 
possible to define the partition function 
\beq
Q_r = \left( \prodr \frac{1}{\ep_{k+1}} \right) \left( 
1 + \sum_{k=1}^{r-1} \de_k + {\cal O}(\nu^2) \right) . 
\lbl{eNQIII} \eeq
This is strongly dependent on the perturbations $\ep_k$, 
implying rapid growth in $Q_r$ with $r$, specifically 
$Q_r={\cal O}(\nu^{-(r-1)})$.   However, the system does 
not approach the equilibrium state $c_r=Q_r$, 
rather it tends to a steady-state configuration. 
In the unperturbed problem this state is $c_r=1$ for all $r$, 
which has the steady-state flux $J=1$; when noise 
is added to the rate coefficients this state is modified to 
$J=1+(\de_1\!-\!\ep_2)+{\cal O}(\nu^2)$ by the calculation 
(\ref{ssflux}).  Since the expected values of $\de_k,\ep_k$ 
satisfy ${\bbf E}[\de_k]=0$ and ${\bbf E}[\ep_k]=\nu>0$, 
the expected value of the steady-state flux is reduced 
by the presence of small amplitude noise from $J=1$ in the noise-free 
system to ${\bbf E}[J]=1-\nu$.   The concentrations asymptote 
to the modified steady-state 
\beq
c_r^{{\rm sss}} = 1 + (\de_1\!-\!\ep_2\!+\!\ep_{r+1}\!-\!\de_r) + 
{\cal O}(\nu^2) . \lbl{eNSSSIII}
\eeq
The presence of ${\cal O}(\nu)$ noise in the reaction rates 
alters the steady-state solution by an amount of ${\cal O}(\nu)$. 
Note that the noise in the two rate constants $a_1,b_2$ affects 
the limiting concentrations of clusters of all sizes.


The kinetics of this case are governed by the 
approach to the steady-state solution (not the 
equilibrium solution). Since the perturbations 
to the fragmentation rates ($\ep_k$) are all positive 
and the perturbations to the coagulation rates can 
be positive or negative, we have ${\bbf E}[\de_k]=0$ 
and ${\bbf E}[\ep_k]=\nu$.   
We replace the differential-difference system by a 
partial differential equation by taking the continuum 
limit and then seeking a diffusive wave solution of this 
partial differential equation.   First, we transform variables 
by $c_r(t) = c_r^{{\rm sss}} \psi_r(t)$ so that $\psi_r(t)
\rightarrow 1$ as $t\rightarrow\infty$ according to 
\beq
\dot \psi_r = 
(1+\de_r)(\psi_{r+1}-\psi_r) + \ep_r(\psi_{r-1}-\psi_r) 
+ \frac{J}{c_r^{{\rm sss}}}(\psi_{r-1}-\psi_{r+1}) , 
\lbl{nkin3eq} \eeq
which formally goes over to the continuum limit 
\beq
\pad{\psi}{t} = \half (1+\de_r+\ep_r) \padd{\psi}{r} - 
(1+\de_r+\ep_r-2\ep_{r+1}) \pad{\psi}{r} .  \lbl{nkin3ypde} 
\eeq
We define $s(t)$ to be the position of the wave and 
transform from  $r$ to an as yet unknown coordinate 
which moves with the diffusive wave by $r=s(t)+z$. 
The wavefront is then determined by $\dot s=1-\nu$, 
an equation which is derived from the leading order 
terms of (\ref{nkin3ypde}) by taking expectation values. 
The speed of propagation is reduced slightly by the 
noisy coefficients, although this is a first-order effect, 
only being present in the ${\cal O}(\nu)$ terms.  The 
large-time and large-size asymptotic solution is 
\beq
c_r(t) \sim \half c_r^{{\rm sss}} \erfc \left( 
\frac{ r - (1\!-\!\nu) t }{\sqrt{2\,(1\!+\!\nu)\,t}} \right) , 
\lbl{eNkin3sol} \eeq
showing that the noise also broadens the diffusive wave 
in aggregation space as in Case I. The erfc shape is 
determined by the higher-order terms of (\ref{nkin3ypde}).


The domination of aggregation over fragmentation in 
the contracted form of Case III is not altered by the 
presence of small amplitude noise 
\beq
\De_n = \sumLn \de_r + {\cal O}(\nu^2) , \hspace*{9mm} 
E_{n+1} = \prodLn \ep_{r+1} . \lbl{eNabIII} 
\eeq
In this case $\alpha_n=1+\De_n$ and the noise in the 
aggregation term, $\De_n$, is ${\cal O}(\nu)$ whereas 
the noise in the fragmentation term, $\beta_{n+1}=E_{n+1}$, is 
much smaller, being of magnitude ${\cal O}(\nu^\lambda)$. 
Superficially this gives the impression of the contracted 
system being more strongly aggregation-dominated than 
the full system; however, this accentuated dominance  
is correct for similar reasons to those expounded for Case I. 
Since ${\bbf E}[\de_r]=0$ and ${\bbf V}[\de_r]={\cal O}(\nu^2)$, 
we have ${\bbf E}[\De_n]=0$ and ${\bbf V}[\De_n]={\cal O}
(\nu^2)$; also ${\bbf E}[\ep_r]=\nu$ and ${\bbf V}[\ep_r] = 
{\cal O}(\nu^2)$ implies ${\bbf E}[E_{n+1}]=\nu^\lambda$ 
and ${\bbf V}[E_{n+1}]={\cal O}(\nu^{2\lambda})$.  
Equations (\ref{URdef}) and (\ref{eNabIII}) imply
\beq
\Upsilon_N \sim \left( \Nprod \frac{1}{E_{n+1}} \right) 
\left( 1 + \Nsum \De_n + {\cal O}(\nu^2) \right) = 
\left( \prod_{r=1}^{\Lambda_N-1} \frac{1}{\ep_{r+1}} 
\right) \left( 1 + \sum_{r=1}^{\Lambda_N-1} \de_r + 
{\cal O}(\nu^2) \right) , 
\eeq
so the contraction procedure does not lose information 
{}from the first two terms of the partition function--to see this, compare 
$\Upsilon_N$ in the above with $Q_R$ in (\ref{eNQIII}) with 
$R=\Lambda_N$. 
However, in this case the partition function does not play 
such an important role as in Cases I and II where it 
determines the large-time asymptotic solution which is 
approached; here, it is the steady-state solution which 
determines the large-time asymptotics.  
Since $E_r = {\cal O}(\nu^\lambda)$, whilst $\De_r = 
{\cal O}(\nu)$, perturbations to the forward coefficients 
influence the first correction term whereas those to the 
backward rates do not.  The steady-state solution is thus 
modified to 
\beq
x_n^{{\rm sss}}  \;=\;  1 + \De_1 - \De_r + {\cal O}(\nu^2) 
\;=\; 1 + \sum_{k=1}^\lambda (\de_k-\de_{(n-1)\lambda+k}) + 
{\cal O}(\nu^2), \lbl{523-sss} 
\eeq
which has constant flux $L=1 + \De_1 + {\cal O}(\nu^2)$.  
The leading-order terms ($x_n=1$, $L=1$) agree with the 
full model and with the model without noisy coefficients, 
however the first correction term is not in agreement with 
the full solution of the noisy model.   The contracted model 
predicts a steady-state flux of $L \sim 1+\sum_{k=1}^\lambda
\de_k + {\cal O}(\nu^2)$, whereas the full model has flux 
$J \sim 1 + \de_1-\ep_2+{\cal O}(\nu^2)$, so agreement is 
limited to the leading order terms ($J=L=1$) only, with the 
first-order correction terms differing, being ${\cal O}(\nu)$.  
In this case microscopic 
detail in the first correction is involved in determining the 
steady-state flux, but the information is lost in the 
coarse-graining contraction, so the procedure only gives 
the correct result to leading order in $\nu$.


To determine the kinetics of approach to steady-state in 
the reduced model we transform from $x_n(t)$ to $\psi_n(t)$ 
by $x_n(t)=x_n^{{\rm sss}}\psi_n(t)$ to gain   
\beq
\dot\psi_n = (1+\De_n)(\psi_{n+1}-\psi_n) + 
E_n(\psi_{n-1}-\psi_n) + \frac{L}{x_n^{{\rm sss}}} 
(\psi_{n-1}-\psi_{n+1}) . \lbl{eNckin3ode} 
\eeq
Since $\De_n={\cal O}(\nu)$ and $E_n={\cal O}(\nu^\lambda)$, 
the two-term continuum expansion of (\ref{eNckin3ode}) 
correct to ${\cal O}(\nu)$ is 
\beq
\pad{\psi}{t} = \half ( 1 + \De_n ) \padd{\psi}{n} - 
( 1 + \De_n ) \pad{\psi}{n} . \lbl{nckin3ypde} 
\eeq
Thus, including the ${\cal O}(\nu)$ terms, the diffusive wave 
has an expected speed of unity since ${\bbf E}[\De_n]=0$. 
This differs slightly from the result for the full model presented 
in (\ref{eNkin3sol}), where the nonzero ${\cal O}(\nu)$ 
perturbations to the fragmentation rates caused the wave 
to be slowed.   After taking expectations, (\ref{nckin3ypde}) 
is solved by $\psi = \half \erfc((n-t)/\sqrt{2t})$ which yields 
the large-time solution
\beq
x_n(t) \sim \half \left( 1 + \De_1 - \De_n \right) \erfc 
\left( \frac{n-t}{\sqrt{2t}} \right) . 
\eeq
Thus coarse-graining has altered the ${\cal O}(\nu)$ 
correction terms in the kinetics of the approach to the 
steady-state, as well as the ${\cal O}(\nu)$ corrections to 
the steady-state itself. However, the leading-order behaviour 
is faithfully reproduced.

\subsection{Case IV}

This generic class of behaviour 
describes systems dominated by fragmentation, in which 
larger clusters break up at a much faster rate than smaller 
ones.


When noise is absent, the partition function is 
identically zero (for $r\geq2$); however, the presence of 
noise alters the partition function to  
\beq
Q_r \sim \frac{1}{[(r-1)!]^p} \left( \prodr \de_k \right) 
\left( 1 - \sum_{k=1}^{r-1} \frac{\ep_{k+1}}{k^p} \right) . 
\lbl{IVnQ} \eeq
Thus $Q_r$ rapidly decays with increasing $r$ since as 
$r\rightarrow\infty$, $Q_r={\cal O}(\nu^{r-1})$ and $p>0$. 
The presence of perturbations to the rate coefficients 
provides a non-zero equilibrium solution $c_r = Q_r$ given 
by equation (\ref{IVnQ}) which is approached, as we shall 
now show, via a diffusive wave.  Transforming to new 
variables $\psi_r(t)=c_r(t)/Q_r$, we find 
\beq
\dot \psi_r=\de_r(\psi_{r+1} -\psi_r)+ \,\left[\, 
(r\!-\!1)^p+\ep_r \,\right]\, (\psi_{r-1}-\psi_r ) ,
\eeq \lbl{eNadvect-diff}
where $\de_r>0$ with ${\bbf E}[\de_r]=\nu$ for all $r$, and 
${\bbf E}[\ep_r]=0$ for all $r$, with $\ep_r$ taking both 
positive and negative values.   In the large $r$ and large time 
limit, $\psi_r$ becomes smooth in $r$ so it is valid to take the 
continuum limit which formally yields 
\beq
\pad{\psi}{t}=\half \,\left[\, (r\!-\!1)^p+\de_r+\ep_r \,\right]\,
\padd{\psi}{r} - \,\left[\, (r\!-\!1)^p+\ep_r-\de_r \,\right]\, 
\pad{\psi}{r} .  \lbl{eNkin4pde}
\eeq
This type of equation has a diffusive wave solution; 
we denote its position by $s(t)$ and transform 
to a coordinate which moves with the wave, via 
$r=s(t)+z$. At leading order, this formally yields the equation 
$\dot s=(s\!-\!1)^p+\ep_s-\de_s$.  Since 
${\bbf E}[\ep_k]=0$ and ${\bbf E}[\de_k]=\nu>0$, 
there is an ${\cal O}(\nu)$ term in the equation for 
the expected value of $s(t)$, namely 
\beq
\dot s=(s\!-\!1)^p-\nu . \lbl{eNkin4seq}
\eeq
Thus the noise has a slowing influence on the wave, 
but since $s\rightarrow\infty$ as $t\rightarrow\infty$, 
the leading-order term is $\dot s_0=s_0^p$, and we have 
\beq
s_0(t) = [(1-p)(t-t_0)]^{1/(1-p)} . \lbl{eNkin4ssol}
\eeq
An ${\cal O}(\nu)$ correction term can be calculated by 
putting $s(t)=s_0(t) + \nu s_1(t)$ into equation (\ref{eNkin4seq}); one 
finds $s_0(t)$ is determined by (\ref{eNkin4ssol}) and $s_1(t)$ 
by $\dot s_1 = -1+ps_0^{p-1} s_1$, which yields 
\beq
s_1 = \frac{-(1-p)(t-t_0)}{(1-2p)} + K (t-t_0)^{p/(1-p)} ,
\lbl{eNkin4s1sol} \eeq
for some constant $K$. 
Thus, for $p<1$, the wave experiences a 
deceleration which reduces the speed by a 
constant amount but which is insufficient to stop the 
wave since the growth in equation (\ref{eNkin4ssol}) has the 
faster growth rate in the limit $t\rightarrow\infty$.  
A higher-order effect is that noise broadens the 
wave very slightly, since noise in (\ref{eNkin4pde}) 
has the effect of increasing the diffusion coefficient. 
The effect of this, however, is minimal since the 
diffusion constant which the wave experiences 
grows without bound in the large time limit. 
In cases where $p<\half$, higher order terms from 
equation (\ref{eNkin4pde}) yield the shape of the front as 
$\psi = \half \erfc(z/\surd(2s(t)+4\nu t))$ so that 
the large-time asymptotic solution (including the 
first correction term due to $\nu$) is given by 
\beq
c_r(t) \sim \half Q_r \erfc \left( \frac{r-s(t)}
{\sqrt{2s(t)+4\nu t}} \right) , \lbl{4shapesol}
\eeq
where $s(t)=s_0(t)+\nu s_1(t)$ is given by equations 
(\ref{eNkin4ssol}) and (\ref{eNkin4s1sol}).  As 
$t\rightarrow+\infty$, the effects of perturbations 
to the rate coefficients decrease in significance, both 
in the position and in the shape of the wavefront.


The rates in the contracted model are given by 
$\alpha_n=\De_n$ and $\beta_{n+1} = n^{p\lambda} 
+ E_{n+1}$ where the perturbations are determined by 
\beq
\De_n = \lambda^{-p\lambda} \prodLn \de_r = 
{\cal O}(\nu^\lambda) , \hspace*{9mm} 
E_{n+1} = n^{p\lambda} \sumLn \frac{ \ep_{r+1} }{r^p} + 
{\cal O}(\nu^2) , \lbl{eNabIV}
\eeq
where in the latter, we have again made the approximation 
(\ref{rgg1app}) valid for large $r$.  Note that, as in Case I, 
$\alpha_r={\cal O}(\nu^\lambda)$ 
whereas $\beta_{r+1}=r^{p\lambda} + {\cal O}(\nu)$,  
suggesting that while the contracted model equations have 
the same structure as the full ones, they are more heavily 
fragmentation dominated, due to the rate at which a cluster 
$x_{n+1}=c_{\Lambda_n+\lambda}$ grows from the cluster 
$x_n = c_{\Lambda_n}$ is much less than for $c_{r+1}$ 
growing from $c_r$. 
If we assume that the perturbations are independent 
and randomly distributed with 
\beq 
{\bbf E}[\de_r]=\nu , \qquad  
{\bbf E}[\ep_{r+1}]=0 , \qquad 
{\bbf V}[\de_r]=V_\delta \nu^2 , \qquad  
{\bbf V}[\ep_{r+1}]=V_\ep\nu^2 , 
\eeq 
then the perturbations to the contracted rates satisfy 
\beq \begin{array}{rclcrcl} 
{\bbf E}[\De_n] & = & \lambda^{-p\lambda}\nu^\lambda, &&  
{\bbf E}[E_{n+1}] & = & 0, \\  
{\bbf V}[\De_n] & = & \lambda^{-2p\lambda} \,[\, 
        (V_\delta\!+\!1)^\lambda\!-\!1\,] \nu^{2\lambda} , &&
{\bbf V}[E_{n+1}] & \approx & V_\ep \lambda^{1-2p} 
        n^{2p(\lambda-1)} \nu^2 ,  
\lbl{514-post-dist} \end{array} \eeq 
where the last formula includes a simplifying approximation 
valid for large $r$ and large $\lambda$, (namely 
replacing $\sum_{r=\Lambda_n}^{\Lambda_{n+1}-1} r^z$ 
with $\int_{r=\Lambda_n}^{\Lambda_{n+1}-1} r^z dr$, 
to yield $\lambda^{1-z} r^{-z}$).  The formulae in 
(\ref{514-post-dist}) show that the contraction procedure 
makes the amplitude of the perturbations to the 
fragmentation rate dependent on the 
aggregation number, with larger sizes having a greater 
variance in the fragmentation rates.   However, at larger sizes 
the standard deviation of the noise grows with 
${\cal O}(n^{p(\lambda-1)})$, thus the noise never 
becomes as large as the deterministic component 
of the rate ($\beta_n={\cal O}(n^{p\lambda})$).


We now use this knowledge of the rates in the contracted 
system including the leading-order perturbations to 
construct a partition function for the contracted system 
$\Upsilon_n$ and, for consistency, verify that $\Upsilon_n 
= Q_r$ when $n=\Lambda_r$. Using (\ref{URdef}) and 
(\ref{eNabIV}), we find 
\beqa
\log \Upsilon_N & \sim & - p\lambda N(\log N-1) 
+ \sum_{n=1}^{N-1} \log \De_n  - \sum_{n=1}^{N-1} 
\frac{E_{n+1}}{n^{p\lambda}}  \nn\\
& \sim & - p\lambda N (\log (\lambda N) -1)
+ \sum_{r=1}^{\Lambda_N-1} \log \de_r 
- \sum_{r=1}^{\Lambda_N-1} \frac{\ep_{r+1}}{r^p} , 
\lbl{524-Up} \eeqa
where a similar approximation to that of (\ref{rgg1app}) 
has been made.   This result should be compared with 
$Q_R$ for $R={\Lambda_N}$ from (\ref{IVnQ}), which yields 
\beq
\log Q_R \sim - p R ( \log R - 1 ) + \sum_{r=1}^{R-1} \log \de_r 
- \sum_{r=1}^{R-1} \frac{\ep_{r+1}}{r^p} . \lbl{524-Q}
\eeq
Thus there are differences between $Q_{\Lambda_N}$ and 
$\Upsilon_N$, but at large $N$ these scale with $\log N$, 
which is of lower order of magnitude than the first two terms 
of $\Upsilon_N$ or $Q_{\Lambda_N}$. For large $N$, the 
leading-order and first correction terms in the log of the 
partition function grow with $N\log N$ and $N$ respectively, 
so both are reproduced correctly in the contracted description, 
as are the perturbations $\de_k,\ep_{k+1}$ as can be 
seen by comparing (\ref{524-Up}) with (\ref{524-Q}); 
it is only high-order correction terms which differ. 
An alternative comparison can be made between 
\beq 
{\bbf E}[\Upsilon_N] = \frac{((N\!-\!1)!)^{p\lambda}}
{ \lambda^{-p\lambda(N-1)} \nu^{\lambda(N-1)}}\qquad 
{\rm and } \qquad {\bbf E}[Q_R] = \frac{((R\!-\!1)!)^p}{\nu^{R-1}},
\eeq
again showing that for large $N$ and $R=\Lambda_N$ 
the leading order terms agree, since Stirling's formula 
leads to the dominant terms in both expressions being 
$(\lambda N)^{p\lambda N} e^{-p\lambda N} / 
\nu^{\lambda N}$.


In the present Case, the system tends to the modified equilibrium 
solution $x_n=\Upsilon_n$ given by (\ref{524-Up}). The 
transformation $\psi_n(t) = x_n(t) / \Upsilon_n$ enables us to 
find the large-time asymptotics; when applied to the equation
\beq
\dot x_n = \De_{n-1}x_{n-1} - \De_nx_n - 
\,\left[\,(n\!-\!1)^{p\lambda}+E_n\,\right]\,x_n 
+ \,\left[\,n^{p\lambda} + E_{n+1}\,\right]\, x_{n+1} ,
\eeq
it yields 
\beq
\dot \psi_n = \De_n ( \psi_{n+1} - \psi_n ) + \,\left[\, 
(n\!-\!1)^{p\lambda} + E_n \,\right]\, (\psi_{n-1}-\psi_n) .
\lbl{534-ode} \eeq
Since $\De_n = {\cal O}(\nu^\lambda)$ whilst $E_n = 
{\cal O}(\nu)$ we keep only those terms involving 
perturbations to the fragmentation rate ($E_n$); thus 
on formally taking the continuum limit, we find 
\beq \pad{\psi}{t}=
\half\,[(n\!-\!1)^{p\lambda}+E_n\,]\,\padd{\psi}{n} - 
[(n-1)^{p\lambda}+E_n\,]\,\pad{\psi}{n}. \lbl{eNckin4pde}
\eeq
The substitution from independent variable $n$ to $z=n-s(t)$ 
yields the expression $\dot s=(s-1)^{p\lambda}$ 
since ${\bbf E}[E_n]=0$; hence the noise has no ${\cal O} 
(\nu)$ effect on the expected speed of the wave in the 
large-time limit unlike the full model (see equation 
(\ref{eNkin4seq})).  Thus as $s\rightarrow\infty$, we have 
$\dot s=s^{p\lambda}$ and so 
$s(t)=[(1\!-\!p\lambda)(t\!-\!t_0)]^{1/(1-p\lambda)}$ 
as in the full model with noiseless rate coefficients.  As in 
Case I, the coarse-graining process has reduced the effect 
of the perturbations. These solutions are only valid for 
$p\lambda<1$; when $p\lambda>1$ the system instantly 
gels as noted by Brilliantov \& Krapivsky \cite{bk}, and 
the large time asymptotics differ significantly. However, 
this case is less relevant physically, since the aggregation 
rate cannot normally grow faster than the cluster size, 
and usually grows much more slowly, giving rise to exponents 
strictly less than unity (i.e.~$p<1$).

\subsection{Case V}

This is perhaps the most interesting of all the nine classes of 
generic behaviour. Both aggregation 
and fragmentation are present and finely balanced; 
both occur faster at larger aggregation numbers than smaller 
ones. As expected in general crystal growth and dissolution processes, 
there is relatively slow 
nucleation of critical nuclei from the free ``monomer'' phase 
and faster growth/dissolution of supercritical clusters.  


In the case with no noise, aggregation and fragmentation 
are perfectly balanced, implying the same value for the 
partition function $Q_r=1$ 
for all $r$.  The presence of perturbations to the rates 
alters this, to
\beq
Q_r = 1 + \sum_{k=1}^{r-1} \left( 
\frac{\de_k-\ep_{k+1}}{k^p} \right) + {\cal O}(\nu^2) . 
\lbl{VnQ} \eeq
Thus for $\nu\ll1$ the leading order behaviour ($Q_r=1$) is 
unaltered, but suffers an ${\cal O}(\nu)$ correction term. 
However, for $p<1$ a result similar to that derived in 
Case II holds, where the effect of the perturbations 
accumulates, so that at large-$r$ the noise may become a 
leading-order effect.  For Case V we assume each 
$\de_k,\ep_k$ has a mean of zero, a variance of $\nu^2$ 
and all are independent random variables.  Thus at large $r$, 
$Q_r$ has an expected value of unity with a variance of 
$\nu^2 r^{1-p}/(1-p)$. 
Thus when $r={\cal O}(\nu^{-2/(1-p)})$, the variance is 
${\cal O}(1)$ and so the perturbations influence the 
leading-order term in the asymptotic expression for $Q_r$. 
Note that if $p>1$ then the variance of $Q_r$ approaches 
$\nu^2\zeta(p)$ (where $\zeta(z)$ is the Riemann zeta function which
arises in the solution of the noiseless time evolution for Case V; see~\cite{rg,rgpap1}), and so the accumulation does 
not become a leading order effect.

In the case $p>1$, however, the system does not evolve 
towards the equilibrium solution, but instead is attracted to a 
steady-state solution with a more rapid decay in the limit 
$r\rightarrow\infty$ which we now examine in more detail. 
Perturbing the rate coefficients modifies this state from one of 
constant flux with $J=1/\zeta(p)$ to $J=1/\zeta(p)+J_1$, where
\beq
J_1 = \frac{1}{\zeta(p)^2} \sum_{k=1}^\infty \left( 
\frac{\de_k}{k^p} \sum_{n=k}^\infty \frac{1}{n^p} - 
\frac{\ep_{k+1}}{k^p} \sum_{n=k+1}^\infty \frac{1}{n^p} \right) . 
\eeq
This gives the steady-state concentrations 
\beq
c_r = \frac{1}{\zeta(p)} \sum_{k=r}^{\infty} \frac{1}{k^p} 
+ \sum_{k=1}^{r-1} \left( \frac{\de_k}{\zeta(p)k^p} 
\sum_{i=k}^\infty \frac{1}{i^p} - \frac{\ep_{k+1}}{\zeta(p)k^p} 
\sum_{i=k+1}^\infty \frac{1}{i^p} - \frac{J_1}{k^p} \right) . 
\lbl{eNSSSV} \eeq 
Unlike the steady-states in Cases III and VI, the perturbations 
$\de_1,\ep_2$ do not play a special role in this solution; 
rather all perturbations influence the steady-state flux 
and concentrations.


Let us now analyse the approach to equilibrium for $p<1$. 
The equilibrium solution is identical to the partition function 
$Q_r$ since $c_1=1$. We transform to new independent 
variables $\psi_r(t)=c_r(t)/Q_r$, with $Q_r$ modified by the 
presence of noise, as in (\ref{VnQ}); for large $r$, we formally 
obtain the continuum approximation 
\beq
\frac{1}{r^p} \pad{\psi}{t} = \left( 1 + \frac{\de_r+\ep_r}{2r^p} 
\right) \padd{\psi}{r} + \left( \frac{p}{r} + 
\frac{\de_r-\ep_r}{r^p} \right)\pad{\psi}{r} .  
\lbl{eNkin5eq} \eeq
This describes physical behaviour that 
approaches equilibrium neither by a diffusive 
wave nor by a simple diffusive process as Case II did; instead 
there is a more complicated similarity solution.  Including 
${\cal O}(\nu)$ correction terms, taking the expectation 
of this leads to $\psi_t=(r^p\psi_r)_r$ so the similarity solution
\beq
\psi(r,t) = \frac{\int_{r/t^{1/(2-p)}}^\infty u^{-p} 
\exp( - u^{2-p} / (2-p)^2) du}
{\int_0^\infty u^{-p} \exp( - u^{2-p} / (2-p)^2) du } ,  
\lbl{5epsi} \eeq
provides the correct asymptotic approximation in the limit 
$\nu\ll1$; the correction terms being much smaller than 
${\cal O}(\nu)$ (that is $o(\nu)$), in contrast with Cases 
IV and VI where the correction terms are ${\cal O}(\nu)$. 

When $p>1$ a steady-state with non-zero flux is approached. 
We convert to new variables $c_r(t) = c_r^{{\rm sss}}\psi_r(t)$.  
For $1<p<2$ the continuum approximation valid at large $r$ 
is formally 
\beq
\frac{1}{r^p} \pad{\psi}{t} = \left( 1 + \frac{\de_r+\ep_r}{2r^p} 
\right) \padd{\psi}{r} + \left( \frac{2-p}{r} + 
\frac{\de_r-\ep_r}{r^p} \right) \pad{\psi}{r} , 
\lbl{eNkinVeq} \eeq
which, when we take the expected value of each term, 
reduces to 
\beq 
\pad{\psi}{t} = r^p \padd{\psi}{r} + (2-p) r^{p-1} \pad{\psi}{r} , 
\lbl{Vcleaneq} \eeq 
the correction terms due to $\delta_r$ and $\ep_r$ again 
being much smaller than ${\cal O}(\nu)$, that is of 
magnitude $o(\nu)$. 
This equation has the solution 
\beq 
\psi = \frac{\int_{r/t^{1/(2-p)}}^\infty u^{p-2} 
\exp(-u^{2-p}/(2-p)^2) \, du}{\int_0^\infty u^{p-2} 
\exp(-u^{2-p}/(2-p)^2) \, du} . 
\lbl{Vcleansol} \eeq 
Because $p>1$, the perturbations are uniformly small in $r$ in 
both advection and diffusion terms. Thus this is a uniformly-valid 
leading-order solution.  For $p\geq2$, the similarity solution 
of equation (\ref{Vcleansol}) is not well-defined, and so does 
not determine convergence to the steady-state.


In the contracted system of equations, the rates are given by 
$\alpha_n = n^{p\lambda} + \De_n$ and $\beta_{n+1} = 
n^{p\lambda} + E_{n+1}$,  where, again using (\ref{rgg1app}), 
we find 
\beq
\De_n = n^{p\lambda} \sumLn \frac{\de_k}{k^p} , \qquad 
E_{n+1} = n^{p\lambda} \sumLn \frac{\ep_{k+1}}{k^p} . 
\lbl{cVnab} \eeq
In contrast to Cases IV and VI, perturbations to the forward 
and backward rates in the contracted equations have the 
same order of magnitude.  If the noise in the full model is 
assumed to be randomly distributed according to 
\beq 
{\bbf E}[\de_r]=0 , \qquad  
{\bbf E}[\ep_{r+1}]=0 , \qquad 
{\bbf V}[\de_r]=V_\delta \nu^2 , \qquad  
{\bbf V}[\ep_{r+1}]=V_\ep\nu^2 , 
\eeq 
then in the contracted model we have 
\beqa 
{\bbf E}[\De_n] =0, \qquad 
{\bbf E}[E_{n+1}]=0, \qquad 
{\bbf V}[\De_n] \approx V_\delta \lambda^{1-2p} 
        n^{2p(\lambda-1)} \nu^2 , &\quad & 
{\bbf V}[E_{n+1}] \approx V_\ep \lambda^{1-2p} 
        n^{2p(\lambda-1)} \nu^2 .  \nn \\ && 
\lbl{515-post-dist} \eeqa 
As in equation (\ref{514-post-dist}), the approximations 
replace $\sum_{r=\Lambda_n}^{\Lambda_{n+1}-1}r^p$ 
by $\int_{r=\Lambda_n}^{\Lambda_{n+1}-1}r^p\,dr$.  From 
(\ref{515-post-dist}) we see that the amplitude of the 
noise is size-dependent in the contracted description 
but, even at large aggregation numbers, it never rivals the 
deterministic component of the rate coefficients.  The 
expected values and the orders of magnitude (in $\nu$) of 
the variance are correctly maintained in the reduced model.


Following a similar method to that of Case IV, we insert the 
contracted rates (\ref{cVnab}) into (\ref{URdef}) to find the 
contracted partition function 
\beq
\Upsilon_N = 1 + \sum_{n=1}^{N-1} 
\frac{\De_n-E_{n+1}}{n^{p\lambda}} + {\cal O}(\nu^2) = 
1 + \sum_{r=1}^{\Lambda_N-1} \frac{\de_r-\ep_{r+1}}{r^p} 
+ {\cal O}(\nu^2) \lbl{eNcQ5}
\eeq
In this case the partition function constructed from the 
reduced rate coefficients matches the first two terms of 
the full (microscopic) partition function $Q_R$ with 
$R=\Lambda_N$ as in (\ref{VnQ}).  This function also provides 
the large-time asymptotic approximation to the equilibrium 
solution for $p\lambda\leq1$.

In cases where $p>1/\lambda$, a steady-state is 
approached instead of the equilibrium. This differs from 
the full model, where the transition from equilibrium to 
steady-state occurs at $p=1$.  Values of $p$ satisfying 
$1/\lambda<p<1$ will approach a steady-state in the 
contracted description but the equilibrium solution in the 
full model. Thus when $0<p<1$, in order to capture the 
correct qualitative behaviour, one must ensure that 
$\lambda<1/p$.

The steady-state solution for $p > 
1/\lambda$ has flux $L=1/\zeta(p\lambda)+L_1$ where 
\beq
L_1 = \frac{1}{\zeta(p\lambda)^2} \sum_{r=1}^\infty 
\left( \frac{\De_r}{r^{p\lambda}} 
\sum_{k=r}^\infty \frac{1}{k^{p\lambda}} - 
\frac{E_{r+1}}{r^{p\lambda}}
\sum_{k=r+1}^\infty \frac{1}{k^{p\lambda}} \right) , 
\lbl{eNcssfV} \eeq
the steady-state being given by 
\beq
x_n^{{\rm sss}} = \frac{1}{\zeta(p\lambda)} \sum_{k=n}^\infty 
\frac{1}{k^{p\lambda}} + \frac{1}{\zeta(p\lambda)}
\sum_{r=n}^\infty \left( 
\frac{L_1 \zeta(p\lambda)}{r^{p\lambda}} + 
\frac{E_{r+1}}{r^{p\lambda}} 
\sum_{k=r+1}^\infty \frac{1}{k^{p\lambda}} -
\frac{\De_r}{r^{p\lambda}} 
\sum_{k=r}^\infty \frac{1}{k^{p\lambda}} \right) . 
\lbl{eNcsssV} \eeq
This clearly has a similar form to (\ref{eNSSSV}); however, 
whilst (\ref{eNcsssV}) displays  the same qualitative 
behaviour as (\ref{eNSSSV}), there are quantitative 
differences.  The leading-order part of (\ref{eNSSSV}) at 
large $r$ decays according to $c_r \sim 
1/(p\!-\!1)\zeta(p)r^{p-1}$, implying that for $r=\Lambda_n$ 
the quantity $c_r$ asymptotes to  $1/ (p\!-\!1) \zeta(p) 
\lambda^{p-1} n^{p-1}$ whilst $x_n \sim 
1/(p\lambda\!-\!1) \zeta(p\lambda) n^{p\lambda-1}$; 
while both decay algebraically, the coarse-graining 
procedure has altered the exponent of the decay.


Let us turn now to the kinetics of the system's approach to 
equilibrium or a steady-state.  Since both $\De_n$ and $E_n$ 
are ${\cal O}(\nu)$ the kinetics of the contracted system 
is very similar to that of the full system.  For $p<1/\lambda$ 
the system approaches the equilibrium 
solution $x_n=\Upsilon_n$, where to ${\cal O}(\nu)$, 
$\Upsilon_n$ is identical to $Q_r$ with $r=\Lambda_n$ 
-- compare equations (\ref{VnQ}) and (\ref{eNcQ5}). 
The manner of approach is remarkably similar to 
(\ref{eNkin5eq}), namely for 
$\psi(n,t)=x_n(t)/\Upsilon_n$ we formally have 
\beq
\frac{1}{n^{p\lambda}}\pad{\psi}{t} = \left( 1 + 
\frac{\De_n+E_n}{2n^{p\lambda}} \right)\padd{\psi}{n} +
\left( \frac{p\lambda}{n} + \frac{\De_n-E_n}
{n^{p\lambda}}\right)\pad{\psi}{n} , 
\lbl{525-epde} \eeq
which, on taking expectations leads to the solution
\beq
\psi(n,t) = \frac
{\int_{n/t^{1/(2-p\lambda)}}^\infty u^{-p\lambda} 
\exp (-u^{-p\lambda} / (p\lambda)^2) \, du}
{\int_0^\infty u^{p\lambda-2} 
\exp (-u^{-p\lambda} / (p\lambda)^2) \, du} . 
\lbl{5sssim} \eeq
When $p>1/\lambda$ the system approaches the 
perturbed steady-state solution $x_n^{{\rm sss}}$ given 
in (\ref{eNcssfV})--(\ref{eNcsssV}) according to 
\beq
\frac{1}{n^{p\lambda}}\pad{\psi}{t} = \left( 1 + 
\frac{\De_n+E_n}{2n^{p\lambda}} \right)\padd{\psi}{n} +
\left( \frac{2-p\lambda}{n} + \frac{\De_n-E_n}
{n^{p\lambda}} \right)\pad{\psi}{n} ,
\lbl{525-spde} \eeq
where $\psi(n,t)=x_n(t)/x_n^{{\rm sss}}$. Equation 
(\ref{525-spde}) corresponds to the kinetics of the full 
model (\ref{eNkinVeq}). As in the full model, there are also 
${\cal O}(n^{-2})$ corrections to the advection term which are 
more significant than the $n^{-p\lambda}$ terms if 
$p>2/\lambda$.   After taking the expectation of equation 
(\ref{525-spde}), the solution 
\beq
\psi(n,t) = \frac
{\int_{n/t^{1/(2-p\lambda)}}^\infty u^{p\lambda-2} 
\exp (-u^{2-p\lambda} / (2-p\lambda)^2) \, du}
{\int_0^\infty u^{p\lambda-2} 
\exp (-u^{2-p\lambda} / (2-p\lambda)^2) \, du} , 
\eeq
can be found.

\subsection{Case VI}

In this generic case there is virtually no fragmentation; the system
is dominated by aggregation which occurs more rapidly for large cluster 
sizes.  This scenario is typical of nucleation and growth processes in 
heavily supersaturated solutions, 
in which critical nuclei form relatively slowly, but then grow 
in size very rapidly.


As in Case III, the partition function is not defined 
when noise is absent since all the fragmentation rates 
are then zero; when the fragmentation rates 
are perturbed by noise the partition function $Q_r$ 
can be defined but is sensitive to the amplitude of $\ep_k$ 
\beq
Q_r \sim [(r-1)!]^p \left( \prod_{k=1}^{r-1} 
\frac{1}{\ep_{k+1}} \right) \left( 1 + \sum_{k=1}^{r-1} 
\frac{\de_k}{k^p} \right) , 
\lbl{VInQ} \eeq
thus $Q_r = {\cal O}(\nu^{-(r-1)})$.   However in this case the 
system does not tend to the equilibrium state $c_r=Q_r$; instead, 
it tends to a steady-state given by constant flux $J = 1 + 
(\de_1-2^{-p}\ep_2)+{\cal O}(\nu^2)$, which implies
\beq
c_r^{{\rm sss}} = \frac{1}{r^p} \left[ 1+ \left(\de_1-2^{-p}\ep_2+
\frac{\ep_{r+1}}{(r\!+\!1)^p} - \frac{\de_r}{r^p} \right) \right] . 
\lbl{VIpsss} \eeq
Thus, altering $a_r,b_{r+1}$ by ${\cal O}(\nu)$ alters the 
steady-state by an ${\cal O}(\nu)$ amount.  As one might 
expect, the perturbations to rates at larger aggregation 
numbers are less significant than those to lower aggregation 
numbers; also as in Case III, the first two perturbations 
$\de_1,\ep_2$ influence all other concentrations at 
${\cal O}(\nu)$.  Whilst the presence of noise restores the 
existence of the partition function, equation (\ref{VIpsss}) is the more 
physically relevant result, in that the partition function which increases 
with $r$ will not be directly manifest in a simulation or experiment, 
where as the steady-state concentrations will be.  These decrease 
with increasing $r$ since $p>0$.


The manner in which the steady-state solution is approached 
is found by substituting $c_r(t) = c_r^{{\rm sss}}\psi_r(t)$ into 
the determining equations (\ref{nkineq}) which, with $b=0$ 
and $a=1$, gives 
\beq
\dot \psi_r =  \ep_r(\psi_{r-1}-\psi_r) + 
(r^p+\de_r)(\psi_{r+1}-\psi_r) +
\frac{J}{c_r^{{\rm sss}}}(\psi_{r-1}-\psi_{r+1}) ,  
\lbl{eNkin6ydot} \eeq
where $J$ is the steady-state flux into the system.  We have 
two-term asymptotic expansions for $J$ and $c_r^{{\rm sss}}$ 
given in, and just before, (\ref{VIpsss}) which enable us to 
simplify this equation.  In the large-time and large aggregation 
number limits, we formally take the continuum limit 
\beq
\pad{\psi}{t} = \half (r^p+\de_r+\ep_r) \padd{\psi}{r} - 
\left( r^p + \de_r + \ep_r - 2 \ep_{r+1} \right) \pad{\psi}{r} . 
\lbl{eNkin6ypde} \eeq
As noted earlier (Case IV, eqn~(\ref{eNadvect-diff})), this 
equation has a solution in the form of 
an advective-diffusive wave; to determine the speed and 
shape of such a wave we transform to a frame of reference 
which moves with the wave by $r=s(t)+z$ where $r=s(t)$ 
denotes the position of the wave.  The leading-order terms 
in (\ref{eNkin6ypde}) yield 
\beq
\dot s = s^p + \de_s + \ep_s - 2\ep_{s+1} , \lbl{6sode}
\eeq
as $s\rightarrow\infty$.  Since ${\bbf E}[\de_k]=0$ and 
${\bbf E}[\ep_k]=\nu>0$, taking expectations leads to 
$\dot s = s^p - \nu$, so perturbations have a small 
slowing effect on the progress of the diffusive wave.  
We solve (\ref{6sode}) by assuming $s(t) = s_0(t)+\nu s_1(t)$ 
with leading order solution  (\ref{eNkin4ssol}), as in Case IV.   
As $s_0\rightarrow\infty$, the ${\cal O}(\nu)$ correction 
term is determined by $\dot s_1=-1+p s_0^{p-1}s_1$ with 
solution (\ref{eNkin4s1sol}).  A higher-order 
effect is that noise broadens the wave very slightly, 
since noise in (\ref{eNkin6ypde}) has the effect of 
increasing the diffusion coefficient. The effect of this, 
however, is minimal since the diffusion constant 
which the wave experiences grows without bound 
in the large time limit. When $p<\half$ the shape 
of the wavefront can be calculated in the same way 
as for Case IV, the solution for $\psi(r,t)$ being identical, 
and leading to 
\beq
c_r(t) \sim \half c_r^{{\rm sss}} \erfc \left( \frac{r-s(t)}
{\sqrt{2s(t)+4\nu t}} \right) ,  
\eeq
where $s(t) = s_0(t) + \nu s_1(t)$ is given by 
eqns~(\ref{eNkin4ssol}) and~(\ref{eNkin4s1sol}).


In this case, coarse-graining maps the rates to 
$\alpha_n=n^{p\lambda}+\De_n$, $\beta_{n+1}=E_{n+1}$ 
where the perturbations to the reaction rates $\De_n,E_{n+1}$
are given by the formulae 
\beq
\De_n= n^{p\lambda} \sumLn \frac{\de_r}{r^p} + 
{\cal O}(\nu^2) ,\hspace*{9mm} 
E_{n+1}=\lambda^{-p\lambda}\prodLn\ep_{r+1} 
= {\cal O}(\nu^\lambda) . \lbl{cVInab} 
\eeq
Thus the domination of aggregation over fragmentation 
persists.   We assume that the perturbations in the full
microscopic model satisfy 
\beq 
{\bbf E}[\de_r] = 0 , \qquad  
{\bbf E}[\ep_{r+1}] = \nu , \qquad 
{\bbf V}[\de_r] = V_\delta \nu^2 , \qquad 
{\bbf V}[\ep_{r+1}] = V_\ep \nu^2 , 
\eeq 
whence we find the perturbations in the reduced description 
satisfy 
\beq 
{\bbf E}[\De_n] = 0, \quad 
{\bbf E}[E_{n+1}]=\lambda^{-p\lambda}\nu^\lambda,\quad 
{\bbf V}[\De_n] \approx V_\delta \lambda^{1-2p} 
        n^{2p(\lambda-1)} \nu^2, \quad 
{\bbf V}[E_{n+1}] = \lambda^{-2p\lambda}  
        [(V_\ep\!+\!1)^\lambda\!-\!1] \nu^{2\lambda} . 
\eeq 
These results are identical to Case IV (\ref{514-post-dist}), 
but with aggregation and fragmentation rates reversed.


Following the method used in the earlier cases, we 
insert (\ref{cVInab}) into (\ref{URdef}) and find 
\beqa
\log \Upsilon_N & \sim & p\lambda N (\log N-1) - 
\sum_{n=1}^{N-1} \log E_{n+1} + 
\sum_{n=1}^{N-1} \frac{\De_k}{k^{p\lambda}} \nn\\ 
& \sim & p\lambda N(\log(\lambda N) - 1 ) - 
\sum_{k=1}^{\Lambda_N-1} \log \ep_{k+1} + 
\sum_{k=1}^{\Lambda_N-1} \frac{\de_k}{k^p} , 
\lbl{eNcQ6} \eeqa
which agrees with a direct expansion of $Q_R$ (\ref{VInQ})
\beq
\log Q_R \sim p R ( \log R - 1 ) - \sum_{k=1}^{R-1} 
\log \ep_{k+1} + \sum_{k=1}^{R-1} \frac{\de_k}{k^p} . 
\eeq
For large $R=\Lambda_N$, the differences between 
$\log Q_R$ and $\log \Upsilon_N$ grow with $\log R$, 
or $\log N$, whereas the quantities themselves grow 
at the much faster rates of $R\log R$ and $N\log N$. 
An alternative comparison can be made between 
\beq 
{\bbf E}[\Upsilon_N] = \frac{ \lambda^{-p\lambda(N-1)} 
\nu^{\lambda(N-1)}}{(N-1)!^{p\lambda}} \qquad {\rm and } 
\qquad {\bbf E}[Q_R] = \frac{\nu^{R-1}}{(R-1)!^p} , 
\eeq
for $R=\Lambda_N$, where Stirling's approximation can be 
used to show that both expressions are dominated by 
$\nu^{\lambda N} (e/\lambda N)^{p\lambda N}$.

However, in this case the system approaches a steady-state 
solution. When the leading order effects of the perturbations 
are incorporated, the steady-state becomes 
\beq
x_n^{{\rm sss}} = \frac{1}{n^{p\lambda}} \,\left[\, 1 + \left( \De_1 -  
n^{-p\lambda} \De_n \right) \,\right] , \lbl{eNcSSS6}
\eeq
which has constant flux $L=1 + \De_1$.   This has the correct 
qualitative behaviour, decaying as $n$ increases. The  
leading-order term for large $n$ is $x_n \sim n^{-p\lambda}$ 
whilst the microscopic model predicts $c_r \sim r^{-p}$ so 
that for maximum accuracy we should have $c_r\sim 
(\lambda n)^{-p\lambda}$ when $r=\Lambda_n$ which 
has a qualitatively similar shape (the decay is algebraic), but 
differs quantitatively since the exponent differs.  The 
steady-state flux also differs. In the full model it is given by 
$J=1+\de_1-\ep_2$ which depends on noise in the 
fragmentation eate coefficient ($\ep_2$) and has the 
expected value ${\bbf E}[J]=1-\nu$ (since ${\bbf E}[\de_r]=0$ 
and ${\bbf E}[\ep_r]=\nu$ for all $r$). In the contracted model 
the flux depends only on the perturbations to the aggregation 
rates, with ${\bbf E}[L]=1+o(\nu)$ since $L=1+\De_1$ 
and $\De_1$ involves the perturbations to the 
aggregation rates $\de_r$ for $r=1,2,\ldots,\lambda$.


To ${\cal O}(\nu)$, the equations which determine 
the kinetics of the approach to steady-state are 
\beq
\dot x_n=\,\left[\,
(n\!-\!1)^{p\lambda}+\De_{n-1}\,\right]\,x_{n-1}-
\,\left[\,n^{p\lambda}+\De_n \,\right]\, x_n , 
\lbl{46eq} \eeq
since $\De_n={\cal O}(\nu)$ and $E_n={\cal O}(\nu^\lambda)$. 
The approach can be elucidated by substituting $x_n(t) = 
x_n^{{\rm sss}}\psi_n(t)$ into (\ref{46eq}), where 
$x_n^{{\rm sss}}$ is as given in (\ref{eNcSSS6}), since then 
\beq
\dot \psi_n = (n^{p\lambda}+\De_n)(\psi_{n+1}-\psi_n) + 
\frac{L}{x_n^{{\rm sss}}}(\psi_{n-1}-\psi_{n+1}).
\eeq
The steady-state solution $x_n^{{\rm sss}}$ has as an asymptotic 
approximation (\ref{eNcSSS6}) with steady flux $L=1+\De_1$.   
In the large-$n$ and large-time limit, we take the continuum 
limit, obtaining the partial differential equation 
\beq
\pad{\psi}{t} = \half (n^{p\lambda}+\De_n) \padd{\psi}{n} - 
(n^{p\lambda}+\De_n) \pad{\psi}{n} .  \lbl{eNckin6pde}
\eeq
Following earlier analysis, we transform to a moving 
coordinate frame by $n=s(t)+z$, obtaining the equation 
$\dot s = s^{p\lambda}+\De_s$ for the position of the 
wavefront. Since ${\bbf E}[\De_s]=0$, there is no 
effect of the noise on the leading-order progression of the 
wave through the system, and we have 
\beq 
s(t)=[(1-p\lambda)(t-t_0)]^{1/(1-p\lambda)} , 
\lbl{536-ssol} \eeq  
the corrections due to noise being of order 
${\cal O}(\nu^2)$ and above. For $p\lambda<\half$ the 
shape of the wave is given by $\psi = \half \erfc ( z 
\sqrt{1-2p\lambda}/\sqrt{2s(t)})$, so that the large-time 
asymptotics are governed by 
\beq
x_n(t) \sim \frac{1}{2n^{p\lambda}} \left( 1 + \De_1 - 
\frac{\De_n}{n^{p\lambda}} \right) \erfc \left( 
\frac{(n-s(t))\sqrt{1-2p\lambda}}{\sqrt{2s(t)}} \right) ,
\eeq
for $n-s(t) \sim \sqrt{s(t)}$ as $t\rightarrow\infty$.   For 
$\half<p\lambda<1$ the shape of the wave depends on 
the initial data for all time.  These results are consistent with 
Cases I, III, and IV, where the coarse-graining procedure 
reduces the effect of perturbations to fragmentation rates.

\subsection{Case VII}

Here fragmentation is the dominant process, with 
smaller clusters shedding monomers much more readily than 
larger clusters.

The coagulation and fragmentation rates are given by 
$a_r = \de_r$ and $b_{r+1} = b r^p +\ep_{r+1}$ with $p<0$, 
and in all of Cases VII, VIII and IX, the perturbations 
are taken to be positive random parameters distributed 
according to 
\beq 
{\bbf E}[\de_r]=\nu , \qquad  
{\bbf E}[\ep_{r+1}]=\nu , \qquad 
{\bbf V}[\de_r]=V_\delta \nu^2 \qquad  
{\bbf V}[\ep_{r+1}]=V_\ep\nu^2 . \lbl{51789-dist}
\eeq 
As with Cases I and IV, if there are no perturbations to the 
rates then the partition function is identically zero; when 
noise is introduced we find 
\beq
Q_r = \prodr \frac{\de_k}{k^p + \ep_{k+1}} , 
\eeq
so that $Q_r$ decreases rapidly in magnitude as $r$ increases 
($Q_r={\cal O}(\nu^{r-1})$). 
However, since $p<0$, this ceases to be valid when 
$r={\cal O}(\nu^{1/p})$, because the perturbations have the 
same order of magnitude as the deterministic part of 
the coefficients. 
When $r\nu^{1/(1-p)}<1$ the approximation (\ref{IVnQ}) is valid; 
however, this ceases to be hold when $r = {\cal O}( 
\nu^{-1/(1-p)})$, where the variance of the sum reaches 
${\cal O}(1)$.  At this point the random perturbations 
affect the leading order behaviour of the partition function.


At large times the system approaches its equilibrium solution. 
As in Case IV, the equation determining the progression 
of the diffusive wave is formally $\dot s=(s-1)^p + \ep_s- 
\de_s$.  On taking expectations of this equation, all 
$\nu$-dependence disappears, indicating that the noise 
has no net effect on the progress of the wave.   The 
equation $\dot s = s^p$ is valid until $s={\cal O}
(\nu^{1/p})$,  hence $s(t)$ follows (\ref{eNkin4ssol}) until 
$t={\cal O}(\nu^{-1+1/p})$.   After this time, the random 
components of the coefficients influence the leading-order 
motion of the wave.   Thus there are subtle differences 
between Case IV ($p>0$), where (\ref{eNkin4ssol}) holds 
for all time, and Case VII ($p<0$), where (\ref{eNkin4ssol}) 
holds only for times up to ${\cal O}(\nu^{1/p})$.
At $r={\cal O}(\nu^{1/p})$ the random component of the 
rate coefficients in the diffusive term in (\ref{eNkin4pde}) 
becomes leading-order. Since this term contains the sum 
of two positive perturbations, there is a net increase in the 
diffusivity, hence the wavefront widens as described by 
(\ref{4shapesol}).


Following the coarse-graining contraction, formulae 
(\ref{eNabIV}) still hold, so the noise satisfies 
\beq \begin{array}{rclcrcl}
{\bbf E}[\De_n] &=& \lambda^{-p\lambda} \nu^{\lambda} ,
&\quad & {\bbf E}[E_{n+1}] & = & n^{p\lambda} \nu 
\ds\sum_{k=\Lambda_n}^{\Lambda_{n+1}-1} \frac{1}{k^p} ,\\  
{\bbf V}[\De_n] & \approx & \lambda^{-2p\lambda} 
\nu^{2\lambda} (V_\delta+1)^\lambda , & \quad &
{\bbf V}[E_{n+1}] & \approx & V_\ep n^{2p\lambda} \nu^2 
\ds\sum_{k=\Lambda_n}^{\Lambda_{n+1}-1} \frac{1}{k^p} . 
\end{array} \lbl{517-post-dist}\eeq
Due to the decay in fragmentation rates at large $r$, there is 
an aggregation size at which the perturbations assume the same 
order of magnitude as the deterministic part of the rate coefficients. 
To find this threshold, we equate $n^{p\lambda}$ with 
the expectation of $E_{n+1}$ in (\ref{517-post-dist}). 
This yields $n=\nu^{1/p}\lambda^{-1+1/p}$, corresponding 
to an aggregation number of $r=\nu^{1/p}\lambda^{1/p}$ 
in the full description of the model (before contraction), 
showing the correct order of magnitude when compared 
with the full model, which gives $r=\nu^{1/p}$.


Using the coarse-grained rate coefficients to construct a partition 
function $\Upsilon_N$, we find (\ref{524-Up}) holds for 
$N={\cal O}(1)$.  Together with equation  (\ref{517-post-dist}) 
and (\ref{IVnQ}), for $R,N={\cal O}(1)$, these imply 
\beq 
{\bbf E}[\Upsilon_N] = \frac{\nu^{\lambda(N-1)}}
{(N\!-\!1)!^{p\lambda}} \left( 1 - \nu \sum_{n=1}^{N-1} 
\frac{1}{n^{p\lambda}} \right) \quad {\rm and} \quad 
{\bbf E}[Q_R] = \frac{\nu^{R-1}}{(R\!-\!1)!^p} \left( 1 - 
\nu \sum_{k=1}^{R-1} \frac{1}{k^p} \right) , 
\lbl{527cf} \eeq
so that for $R=\Lambda_N$ we have both expressions 
growing with $R$ (or equivalently with $N$), 
whilst their difference only grows with $\log R$.  
This shows nice agreement. 
At large $R,N$ the formulae (\ref{527cf}) fail due to the 
noise in the fragmentation rate becoming leading order. 
This occurs for $R = {\cal O}(\nu^{1/(p-1)})$ in the latter case 
and, upon using (\ref{517-post-dist}), $N = 
{\cal O}(\nu^{1/(p-1)})$ in the former; again showing good 
agreement between the full and contracted models.


As with Case IV, the system converges to equilibrium via a 
diffusive wave which travels from small to large-$n$, invading 
the region where the initial conditions $x_n=0$ have not yet 
been modified, and leaving behind (at smaller $n$) the 
equilibrium solution $x_n=\Upsilon_n$.  Thus we use the 
substitution $\psi_n(t) = x_n(t)/\Upsilon_n$, which satisfies 
(\ref{534-ode}).   Since ${\bbf E}[\De_n] = 
{\cal O}(\nu^\lambda)$ and ${\bbf E}[E_{n+1}] \sim \lambda^{1-p} 
n^{p(\lambda-1)} \nu$ from (\ref{517-post-dist}),  the 
continuum limit of (\ref{534-ode}) to ${\cal O}(\nu)$ reduces to 
\beq
\pad{\psi}{t} = \left( (n\!-\!1)^p\lambda + 
\lambda^{1-p} n^{p(\lambda-1)} \nu \right) \left( 
\frac{1}{2} \padd{\psi}{n} - \pad{\psi}{n} \right) . 
\eeq
At large-times, the leading-order term determines the speed 
of the wavefront ($\dot s$), giving the equation $\dot s = 
s^{p\lambda}+\lambda^{1-p} s^{p(\lambda-1)} \nu$. 
As $\nu\rightarrow0$, the solution of this equation can be 
formulated as $s(t) = s_0(t)+\nu s_1(t)$ where 
\beq
s_0(t) = \,[\,(1-p\lambda) t \,]^{1/(1-p\lambda)} , 
\quad {\rm and} \quad 
s(t) \sim s_0(t) \left( 1 + \frac{\nu \lambda^{1-p}s_0(t)^{1-p}}
{2-p-p\lambda} \right) . 
\eeq
This shows that as the wave reaches larger values of $n$, 
perturbations to the rate coefficients influence the {\em 
leading-order} motion of the wave.  This occurs when $s\sim 
n={\cal O}(\nu^{1/(p-1)})$, or when $t={\cal O}(\nu^{-1
-p(\lambda-1)/(p-1)})$, which is the same order of 
magnitude as in the uncontracted case.  At this order of 
magnitude the perturbations to the rate coefficients also cause the shape 
of the wavefront to be modified.

\subsection{Case VIII}

Here aggregation and 
fragmentation are finely balanced, but occur much more rapidly 
at small cluster sizes than at large cluster sizes, a situation that
may arise in growth processes in solution 
below saturation without nucleation barriers.


In the noise-free case this system converges to the equilibrium 
solution $c_r=1$. When noisy coefficients are introduced, 
defined by $a_r = a r^p + \de_r$, $b_{r+1}=b r^p + \ep_{r+1}$, 
with $p<0$ and the perturbations distributed according to 
(\ref{51789-dist}), then the partition function is modified to  
\beq 
Q_r = \prodr\frac{1+k^{-p}\de_k}{1+k^{-p}\ep_{k+1}} , 
\eeq 
and so for $r={\cal O}(1)$, the equilibrium 
solution can be approximated by 
\beq
c_r = 1 + \sum_{k=1}^{r-1} k^{-p} (\de_k-\ep_{k+1} ) + 
{\cal O}(\nu^2) . \lbl{8crsss}
\eeq
However, this approximation ceases to be valid at large $r$ 
due to the perturbations becoming a leading-order effect. 
Given that the variance of each perturbation is 
${\cal O}(\nu^2)$, the variance of $c_r$ in equation~(\ref{8crsss}) 
is ${\cal O}(\nu^2 r^{1-p})$ so that at $r=\nu^{-2/(1-p)}$, 
the leading order expression for $c_r$ is no longer unity.  At this 
aggregation number the cumulative effect of the perturbations 
becomes as important as the leading-order term.


When we turn to the analysis of the kinetics of the approach 
to equilibrium, we note the similarities between this 
case (where $p<0$), and the approach to equilibrium in 
Case V for $0<p<1$.   The continuum equation (\ref{eNkin5eq}) 
is also valid here, 
but whereas in Case V ${\bbf E}[\de_r+\ep_r]=0$, here 
${\bbf E}[\de_r+\ep_r]=2\nu$, so when we take the 
expectation of (\ref{eNkin5eq}) we obtain an equation 
with an increased diffusivity, namely 
\beq
\pad{\psi}{t}=(r^p+\nu)\padd{\psi}{r}+pr^{p-1}\pad{\psi}{r}.
\eeq
Thus there is a large-$r$ region where the perturbations 
influence the leading-order kinetics. This occurs for $r = 
{\cal O}(\nu^{1/p})$ and larger, where noise in the diffusion 
term is comparable with the deterministic part of the diffusivity. 
In the advection term, the noise has zero mean.  Thus for 
$r \nu^{-1/p}<1$, we expect the similarity solution $\psi=f(\eta)$ 
with $\eta=r/t^{1/(2-p)}$ to be valid and $\psi(r,t)$ determined 
by equation (\ref{5epsi}); but for $r\geq{\cal O}(\nu^{1/p})$, 
the leading-order kinetics are influenced by the noise.


In the coarse-grained contracted system, the forward and 
backward rate coefficients are given by 
\beq
\alpha_n = \prodLn (r^p + \de_r ) , \hspace*{9mm} 
\beta_{n+1} = \prodLn (r^p + \ep_{r+1} ) , 
\eeq
and whilst (\ref{cVnab}) is a valid approximation for small $n$, 
it breaks down at larger $n$.  Assuming that the noise is 
distributed according to (\ref{51789-dist}) before the 
contraction procedure, we find 
\beqa
{\bbf E}[\De_n] = {\bbf E}[E_{n+1}] = \nu \lambda^{1-p} 
n^{p(\lambda-1)}, &\;\; & 
{\bbf V}[\De_n] = V_\delta n^{2p\lambda} \nu^2 
\sum_{k=\Lambda_n}^{\lambda_{n+1}-1} \frac{1}{k^{2p}} , 
\quad  {\bbf V}[E_{n+1}] = V_\ep n^{2p\lambda} \nu^2 
\sum_{k=\Lambda_n}^{\Lambda_{n+1}-1} \frac{1}{k^{2p}} , 
\nn\\ && \lbl{eNab8exp} \eeqa
following the coarse-grained renormalisation.  The even balance of 
aggregation and fragmentation is reflected in the identical 
expectations of $\De_n$ and $E_{n+1}$ and in the same order 
of magnitude of their variances.  When $n={\cal O}(\nu^{1/p})$ 
the formulaic component of the rates 
(${\cal O}(\nu^{p\lambda})$) matches the random element 
(providing $\lambda$ is not large); this aggregation number 
agrees with the analysis of the fully microscopic model.


We now use the coarse-grained rates to calculate the partition 
function ($\Upsilon_n$) for the coarse-grained system. 
Equation (\ref{eNcQ5}) is valid for $r={\cal O}(1)$, 
however, since $p$ is negative the sums are 
divergent at large $n$ and large $r$.  Although 
${\bbf E}[\Upsilon_N]=1={\bbf E}[Q_R]$ for all $N$ and
for all $R$, the variance of $\Upsilon_N$ grows with $N$ 
according to 
\beq
{\bbf V}[\Upsilon_N] =  \nu^2 (V_\delta+V_\epsilon) 
\sum_{n=1}^{(N-1)\lambda+1} \frac{1}{n^{2p}} , 
\eeq
which agrees with ${\bbf V}[Q_R]$ for $R=\Lambda_n$. 
The asymptotic expression (\ref{eNcQ5}) thus ceases to be 
uniformly valid when the variance becomes ${\cal O}(1)$, 
namely when $n={\cal O}(\nu^{-2/(1-p)})$ giving the 
same order of magnitude of aggregation number 
($r=\Lambda_n$) as can be derived from the full system 
of equations (\ref{VnQ}).


The kinetics of approach to equilibrium are similar to Case V 
with $p<1/\lambda$; that is by a perturbed similarity solution
rather than a moving diffusive wavefront. Since in the present case the 
perturbations obey equation (\ref{51789-dist}), when we take 
the expectation of equation (\ref{525-epde}) we obtain 
\beq
n^{-p\lambda} \pad{\psi}{t} = \left( 1 + 
2 \lambda^{1-p} n^{-p} \nu \right) \padd{\psi}{n} + 
\frac{p\lambda}{n} \pad{\psi}{n} . 
\eeq
Thus when one ignores the ${\cal O}(\nu)$ terms, a similarity 
solution can be found; and for $n<{\cal O}(\nu^{1/p})$, we 
expect the similarity solution to give the leading order 
behaviour.  This solution has the form $\psi_n(t)=f(\eta)$ with 
$\eta=n/t^{1/(2-p\lambda)}$ as in (\ref{5sssim}).  However, 
for $n={\cal O}(\nu^{1/p})$ and larger, another more 
complicated solution takes effect, where the noise influences 
the dynamical behaviour at leading order. For $n$ larger than 
${\cal O}(\nu^{1/p})$ the system behaves as if all rates had 
been chosen at random.

\subsection{Case IX}

The final case corresponds to a system in which 
aggregation dominates fragmentation at smaller cluster sizes; 
at larger cluster sizes, the perturbations to both aggregation 
and fragmentation rate coefficients entirely swamp this effect.

Here, we have $a_r = r^p + \de_r$, $b_{r+1}=\ep_{r+1}$ with the 
perturbations $\de_r,\ep_{r+1}$ distributed according to 
(\ref{51789-dist}). 
Formally the presence of noise allows the partition 
function to be written as 
\beq
Q_r = \prodr\frac{k^p+\de_k}{\ep_{k+1}} . 
\eeq
With noiseless rate coefficients, the system evolves to the 
steady-state with unit flux ($J=1$); however, using the perturbed 
coefficients in equation (\ref{gsss}), 
a calculation of the steady-state 
flux leads to $J=0$ since at large values of $r$, the term $1/ a_r 
Q_r$ has a positive expectation value which is independent of 
$r$, making the sum divergent.  Thus in this case the presence of 
noisy coefficients changes the large-time behaviour from an
approach to steady-state with unit flux to an approach to the 
equilibrium solution, which is given by $c_r=Q_r$ where, for 
$r={\cal O}(1)$, $Q_r$ is given by equation (\ref{VInQ}).   
This expression, however, ceases to be valid for 
values of $r$ larger than ${\cal O}(\nu^{-1/(1-p)})$. 
At intermediate times, before mass has been transported 
to larger aggregation numbers, we expect the steady-state
solution to be manifest. 
At larger times, the system undergoes a complex transition 
from a steady-state solution to the equilibrium solution. 
Evolution will then be dominated by the random perturbations 
to the rate coefficients, and will follow the kinetics described 
in the next section.


In the coarse-grained description of the system 
the equation for $\De_n$ in (\ref{cVInab}) is valid for small $n$, 
but ceases to be valid when $n = {\cal O}(\nu^{1/p})$, 
where the noise becomes leading order;  thus there is an 
intermediate asymptotic regime where the system remains 
aggregation-dominated. Using (\ref{51789-dist}) 
we find the distribution of perturbations to be as follows:
\beq \begin{array}{rclcrcl} 
{\bbf E}[\De_n] & = & n^{p\lambda} \nu 
\ds\sum_{k=\Lambda_n}^{\Lambda_{n+1}-1} \frac{1}{k^p},
&\quad&{\bbf E}[E_{n+1}] & = & \lambda^{-p\lambda} 
\nu^\lambda , \\ 
{\bbf V}[\De_n] & = & V_\de n^{2p\lambda}\nu^2 
\ds\sum_{k=\Lambda_n}^{\Lambda_{n+1}-1} \frac{1}{k^{2p}},
&\quad&
{\bbf V}[E_{n+1}] & = & \lambda^{-2p\lambda} \nu^{2\lambda} 
\,\left[\, (V_\ep+1)^\lambda - 1 \,\right]\, ; 
\end{array} \eeq 
however, at larger aggregation numbers the system 
is dominated by the random perturbations.


Calculating the partition function $\Upsilon_N$ from the 
coarse-grained rates we find  (\ref{eNcQ6}) 
holds for small and intermediate values of $N$. For large $N$ 
however it fails since $p$ is negative and the second 
term becomes the same order of magnitude as the 
first when $N={\cal O}(\nu^{1/p})$.   The presence of 
noise transforms this case from one which approached 
steady-state solution (with $L\approx1$) to one which 
approaches equilibrium (i.e. $L=0$); as in the microscopic 
system.  Thus the coarse-graining procedure has 
faithfully retained the structural difference that noise has 
made to the full system of equations.


In the large-time limit, this case shows convergence to the 
equilibrium solution, since at large aggregation numbers, 
the system appears identical to the case with all rates chosen 
at random described below, in Section \ref{CaseX}.  In this 
case the kinetics are dominated by the random perturbations 
to the rates, so it is impossible to give a detailed description 
of the time-dependent solution.

At intermediate times, when there is little mass in large 
aggregation numbers, however, we expect the system to 
behave as in Case VI, namely the approach to a steady-state 
solution.  Since the rate coefficients at small and intermediate 
aggregation numbers are dominated by aggregation, we 
expect a diffusive wave to move into the larger-$n$ region 
leaving behind the steady-state solution $x_n = 
n^{-p\lambda}$.  Since this moves at a rate given by 
(\ref{536-ssol}), when $t={\cal O} (\nu^{1/p-\lambda})$ 
the wave reaches aggregation sizes $n = {\cal O} 
(\nu^{1/p})$, where the perturbations to the rates are of the 
same order of magnitude as the deterministic 
component of the rates.  Thus after this time the 
system undergoes a transition from the state 
\beq
x_n(t) \sim \left\{ \begin{array}{lcl} 
n^{-p\lambda} && {\rm for} \;\; n \ll \nu^{1/p} \\ 
0 && {\rm for} \;\; n \gg \nu^{1/p} , \end{array} \right. 
\eeq
to the equilibrium state $x_n=\Upsilon_n$.

\subsection{The system with totally random rates} 
\slbl{CaseX}

The last three subsections show that, at large times 
and with the rate perturbations as defined in Section 
\ref{noise-def-sec}, in each of the Cases VII, VIII and 
IX the temporal evolution tends to the 
equilibrium solution rather than a steady-state solution; 
at large $r$ the partition function $Q_r$ is 
dominated by the perturbations, hence so too is the equilibrium 
solution in all three cases.  From this perspective, these 
three cases can 
then be thought of as lying in the same universality class, 
namely a  \BD\ system with ``totally random'' rate coefficients. By 
this we mean a formulation of the \BD\ equations (\ref{fullBD}) 
in which all the rate coefficients $a_r,b_r$ are chosen at random 
(but remain independent of time); in the notation of this 
paper, we have $a=b=0$, leaving $a_r=\de_r$, $b_r=\ep_r$. 
Here the rates $\de_r$, $\ep_r$ are independent random 
parameters whose expectation and variance is size-independent.

If we take the monomer concentration to be 
fixed at unity, the governing equations are  
\beq
\dot c_r = \de_{r-1} c_{r-1} - \ep_r c_r - \de_r c_r 
+ \ep_{r+1} c_{r+1} , \quad r\geq 2 . \lbl{randbd}
\eeq
It is possible to define a partition function $Q_r$ from 
the coefficients $\de_r,\ep_r$ provided none are zero. 
If we assume that in the large-time limit the system 
will tend to a steady-state 
\beq
c_r = Q_r \left( 1 - J \sum_{k=1}^{r-1} \frac{1}{\de_kQ_k} 
\right) , 
\eeq
then imposing the fastest possible decay on the 
concentrations in the limit of large $r$, we find 
\beq
\frac{1}{J} = \sum_{k=1}^\infty \frac{1}{\de_kQ_k} . 
\eeq
Since neither $Q_k\rightarrow\infty$ or $\delta_k 
\rightarrow\infty$ as $k\rightarrow\infty$ 
the terms in the sum do not decay, so the sum diverges. 
We thus have $J=0$, and the system must tend to the 
equilibrium solution $c_r=Q_r$. To find the kinetics in the 
large-time limit we substitute $c_r = Q_r \psi_r$; then $\psi_r$
satisifes the difference equation
\beq
\dot \psi_r = \ep_r\psi_{r-1} - \de_r\psi_r - \ep_r\psi_r + 
\de_r\psi_{r+1} ,
\eeq
This is a system of equations similar to that with which 
we started out in equation (\ref{randbd}); the essential 
difference is that $\psi\rightarrow1$ as $t\rightarrow\infty$. 
We expect $\psi_r(t)$ to be slowly varying in the 
large-time limit, in turn enabling us to take 
the continuum limit which gives
the following partial differential equation for $\psi_r$:
\beq
\pad{\psi}{t} = \half(\de_r+\ep_r) \padd{\psi}{r} + 
( \de_r-\ep_r ) \pad{\psi}{r} . 
\eeq
Thus the advective component of the process 
$(\de_r-\ep_r)\pad{\psi}{r}$ has zero mean (since 
${\bbf E}[\de_r] = {\bbf E}[\ep_r] = \nu$), whilst the diffusive 
component is always positive, having mean $\nu$. The 
expectation of the solution is thus $\psi = \half \erfc
(r/2\sqrt{\nu t})$ and so we see that the system shares 
several similarities 
with Case II, notably convergence to equilibrium via the 
same kinetics albeit on a slower timescale.

In Cases VII and IX we expect the waves to cease 
moving when they reach the large-$r$ region where 
perturbations become dominant, namely when 
the wave front reaches $r = {\cal O}(\nu^{1/p})$; which 
occurs after a time of $t=1/(1-p)\nu^{(1-p)/p^2}$. 
However, in all three Cases VII-IX the final approach to equilibrium 
is by a predominantly diffusive mechanism.

\section{Discussion} \lbl{discussion}
\setcounter{equation}{0}

A previous paper described in detail nine generic classes
of behaviour into which the asymptotic dynamics of the
\BD\ equations with power-law coefficients falls \cite{rgpap1}. 
The nine classes capture 
qualitatively different physical properties which are 
shared by all models within the same class. 
In the present paper, we have concentrated on 
the detailed analysis of these nine cases;  for each case 
we have considered the effect of the perturbations 
to the rate constants on the equilibrium or steady-state 
solution and on the large-time asymptotics.

In Cases IV--VI, where the rate constants grow with increasing 
cluster size, larger perturbations could be considered, 
(that is the perturbations to the forward and backward rate
coefficients $\delta_r,\ep_r$ are each ${\cal O}(1)$). Because the 
unperturbed 
rates grow with $r$, there would then be a large-$r$ region 
where the perturbations are insignificant and a smaller-$r$ 
region where they should be taken into account.  Since we 
are dealing with a discrete system, the small-$r$ region 
consists of a complicated finite dimensional system joined 
to a simpler large-$r$ system whose behaviour we already 
know by asymptotic analysis (King \& Wattis \cite{wk99}). 
By taking a coarse-graining contraction with $\lambda$ 
sufficiently large,  the whole of this small $r$ regime could be 
mapped to a small dimensional system with only a few 
concentrations needing to be retained. 
As in the unperturbed problem analysed previously~\cite{rg,rgpap1}, 
coarse-graining retains the correct leading-order structure 
of the problem, although critical exponents of $p=1$ where the 
behaviour changes are mapped to $p=1/\lambda$.

Cases VII--IX are much more complex, since here the rate 
coefficients decay with increasing aggregation number. 
Thus in the limit $r\rightarrow\infty$, the perturbations 
$\de_r,\ep_r$ dominate the partition function ($Q_r$) 
and the large-time kinetics.  If we assume the perturbations 
are small (as we have done here),  there is a small-$r$ 
region where regular behaviour occurs, and a larger-$r$ 
region where perturbations dominate the rate coefficients, 
the partition function and the kinetics; in this region the 
system behaves as if the rate coefficients had been 
chosen at random.  Following an application of the 
coarse-graining process the positions of these regions 
remain invariant and, in the small $r$ region, the leading 
order behaviour agrees with that of the full model.
The only case in which the large time solution which is 
approached is altered by the presence of noise is 
Case IX, which changes from convergence to a 
steady-state solution in the large-time limit into a 
system which converges to equilibrium.  The equilibrium 
solution to which Cases VII and VIII approach in the 
large-time limit also suffer major modification due to 
the presence of noise.

In renormalisation-theoretic jargon, Cases I--VI can be 
referred to as {\em universality classes}, whereas this is 
not in general true for Cases VII--IX, because the 
perturbations destabilise the steady-state solution, 
modify the equilibrium solution and alter the 
associated large-time dynamics of convergence to 
equilibrium.  However all of Cases VII-IX share 
the same large aggregation number and large time 
behaviour, and so fall into a seventh Universality class, 
which we have referred to as the system with 
totally random rates (Section \ref{CaseX}).
This classification occurs with the rate perturbations 
as described in Section \ref{noise-def-sec}, namely 
those with size-independent mean and variance.

If an alternative form of rate-perturbations were applied 
to the system, then an alternative universality classification 
would be produced.   For example, if the rate-perturbations 
had the same size-dependence as the power-law 
component, that is ${\bbf E}[\de_r] = \nu r^p$ and 
${\bbf E}[\ep_{r+1}]=\nu r^p$ then the large-time 
asymptotics of Cases VII-IX suffer no leading-order 
modification due to the perturbations and each of these 
cases then corresponds to its own universality class. 
However, this one-to-one correspondance of our 
nine special cases with universality classes only holds 
in situations where the added noise decays faster 
than the specified power-law for the rate-coefficients 
as $r\rightarrow\infty$.  We believe such scenarios to be 
somewhat artificial, thus in this paper we have concentrated 
our analysis on the more generic and interesting case 
where the behaviour of the rate-perturbations at large 
aggregation numbers differs from that of the power-law.

In future work, we plan to extend and generalise 
these renormalisation-theoretic 
results to a range of other \BD\ systems, including ones with
different forms of 
rate coefficients which describe 
quite distinct physicochemical 
processes~\cite{expo}, as well as the constant-density 
formulation of the \BD\ equations.  It turns out that, for 
certain well motivated choices of 
rate coefficients, the renormalisation procedure
is {\em exact} (in the sense that the 
approximation made after equation (\ref{rgg1app}) is not 
necessary).  The constant density \BD\ system has the 
added complication of being inherently nonlinear, and 
our renormalisation scheme accentuates the nonlinearity, 
making analysis of this problem even more challenging.

\subsection*{Acknowledgements}

We are grateful to John Cardy for several useful discussions 
regarding renormalisation theory, and to Bob O'Malley for 
pointing out related work in the literature.   PVC is grateful 
to the Department of Theoretical Physics and Wolfson College, 
University of Oxford, for a Wolfson-Harwell Visiting Fellowship, 
and for hosting our discussions.  JADW thanks John King 
for many instructive conversations.


\end{document}